\documentclass[aps,prc,twocolumn,showpacs,floatfix,nobibnotes,nofootinbib,superscriptaddress,amsmath,amssymb,longbibliography]{revtex4-1}

\usepackage{graphicx}
\usepackage{epsfig}
\usepackage{bm}
\usepackage{color}
\usepackage{float}
\usepackage{dcolumn}
\usepackage{multirow} 
\usepackage{hyperref}
\usepackage{placeins}
\usepackage[caption=false]{subfig}

\graphicspath{{.}{../figs/}}

\newcommand{\beq}{\begin{equation}}
\newcommand{\eeq}{\end{equation}}
\newcommand{\beqa}{\begin{eqnarray}}
\newcommand{\eeqa}{\end{eqnarray}}

\newcommand{\DNNLO}{$\Delta$NNLO}

\newcommand{\xEFT}{$\chi$EFT}

\begin{document}
\title{$\Delta$ isobars and nuclear saturation}
\thanks{This manuscript has been authored by UT-Battelle, LLC under
  Contract No. DE-AC05-00OR22725 with the U.S. Department of
  Energy. The United States Government retains and the publisher, by
  accepting the article for publication, acknowledges that the United
  States Government retains a non-exclusive, paid-up, irrevocable,
  world-wide license to publish or reproduce the published form of
  this manuscript, or allow others to do so, for United States
  Government purposes. The Department of Energy will provide public
  access to these results of federally sponsored research in
  accordance with the DOE Public Access
  Plan. (http://energy.gov/downloads/doe-public-access-plan).}

\author{A.~Ekstr\"om} \affiliation{Department of Physics,
  Chalmers University of Technology, SE-412 96 G\"oteborg, Sweden}

\author{G.~Hagen} \affiliation{Physics Division, Oak Ridge National
  Laboratory, Oak Ridge, TN 37831, USA} \affiliation{Department of
  Physics and Astronomy, University of Tennessee, Knoxville, TN 37996,
  USA}

\author{T.~D.~Morris} \affiliation{Department
  of Physics and Astronomy, University of Tennessee, Knoxville, TN
  37996, USA} \affiliation{Physics Division, Oak Ridge National
  Laboratory, Oak Ridge, TN 37831, USA} 

\author{T.~Papenbrock} \affiliation{Department
  of Physics and Astronomy, University of Tennessee, Knoxville, TN
  37996, USA} \affiliation{Physics Division, Oak Ridge National
  Laboratory, Oak Ridge, TN 37831, USA} 

\author{P.~D.~Schwartz} \affiliation{Department
  of Physics and Astronomy, University of Tennessee, Knoxville, TN
  37996, USA} \affiliation{Physics Division, Oak Ridge National
  Laboratory, Oak Ridge, TN 37831, USA}

\begin{abstract} 
  We construct a nuclear interaction in chiral effective field theory
  with explicit inclusion of the $\Delta$-isobar $\Delta(1232)$ degree of
  freedom at all orders up to next-to-next-to-leading order (NNLO). We
  use pion-nucleon ($\pi N$) low-energy constants (LECs) from a
  Roy-Steiner analysis of $\pi N$ scattering data, optimize the LECs
  in the contact potentials up to NNLO to reproduce low-energy
  nucleon-nucleon scattering phase shifts, and constrain the
  three-nucleon interaction at NNLO to reproduce the binding energy
  and point-proton radius of $^{4}$He. For heavier nuclei we use the
  coupled-cluster method to compute binding energies, radii, and
  neutron skins.  We find that radii and binding energies are much
  improved for interactions with explict inclusion of $\Delta(1232)$,
  while $\Delta$-less interactions produce nuclei that are not bound
  with respect to breakup into $\alpha$ particles. The saturation of
  nuclear matter is significantly improved, and its symmetry energy is
  consistent with empirical estimates.
\end{abstract}


\maketitle

\section{Introduction}
In recent years, {\it ab initio} calculation of atomic nuclei with
predictive power have advanced from
light~\cite{kamada2001,pieper2001,navratil2009,barrett2013} to
medium-mass
nuclei~\cite{hagen2014,lahde2014,hagen2015,hergert2016,hagen2016b}.
Such calculations are only as good as their input,
i.e. nucleon-nucleon ($NN$) and three-nucleon ($NNN$) interactions,
therefore the quest for more accurate and more precise nuclear
potentials is an ongoing endeavor at the forefront of
research~\cite{epelbaum2000,epelbaum2002,entem2003,shirokov2004,hebeler2011,ekstrom2013,entem2015,epelbaum2015,ekstrom2015a,lynn2016,carlsson2016}.
Here, potentials from chiral effective field theory (\xEFT) -- based
on long-ranged pion exchanges and short-ranged contact interactions --
play a dominant role~\cite{epelbaum2009,machleidt2011}, because they
are expected to deliver accuracy (via fit to data) and precision (via
increasingly higher orders in the power counting).  As it turns out,
however, state-of-the-art \xEFT~potentials that are accurate for the
lightest nuclei with masses $A=2,3$ vary considerably in their
saturation point for nuclear matter~\cite{hebeler2011} and in their
binding energy for heavier nuclei~\cite{binder2013b,
  carlsson2016,simonis2017}.

This sensitivity of the saturation point to the details of the
\xEFT~interaction is not well understood~\cite{elhatisari2016} and
also puzzling from an EFT perspective. A practical approach to this
dilemma consists of constraining \xEFT~potentials to reproduce
experimentally determined binding energies and charge radii of nuclei
as heavy as oxygen~\cite{ekstrom2015a}. In this work, we will follow a
different approach and explicitly include the $\Delta$ isobar
$\Delta(1232)$, abbreviated $\Delta$ in the following, as a low-energy
degree of freedom in addition to pions ($\pi$) and nucleons ($N$). We
recall that the $\Delta$-$N$ mass-splitting $\delta \equiv
M_{\Delta}-M_{N}\approx$~293~MeV is roughly twice the pion mass ($M_{\pi}
\sim 140$~MeV) and well below the expected breakdown scale of
\xEFT~potentials~\cite{epelbaum2009,machleidt2011}. Furthermore, the
$\Delta$ also couples strongly to the $\pi N$ system. For these
reasons, the early chiral $NN$
interactions~\cite{vankolck1994,ordonez1994,ordonez1996} included the
$\Delta$ degree of freedom. Indeed, \citeauthor{vankolck1994} as well
as \citeauthor{Bernard1997} showed that the low-energy constants
(LECs) of the $\pi N$ interaction in a $\Delta$-less \xEFT{} receive
a substantial contribution via resonance saturation. As nuclear
interactions from {\xEFT} with and without $\Delta$'s have a similar
structure otherwise, only little effort was invested in
producing quantitative $\Delta$-full \xEFT~potentials.  We refer the
reader to the reviews~\cite{epelbaum2009,machleidt2011} for extensive
discussions of this topic.

\begin{figure}[htb]
\begin{center}
\includegraphics[width=0.45\textwidth]{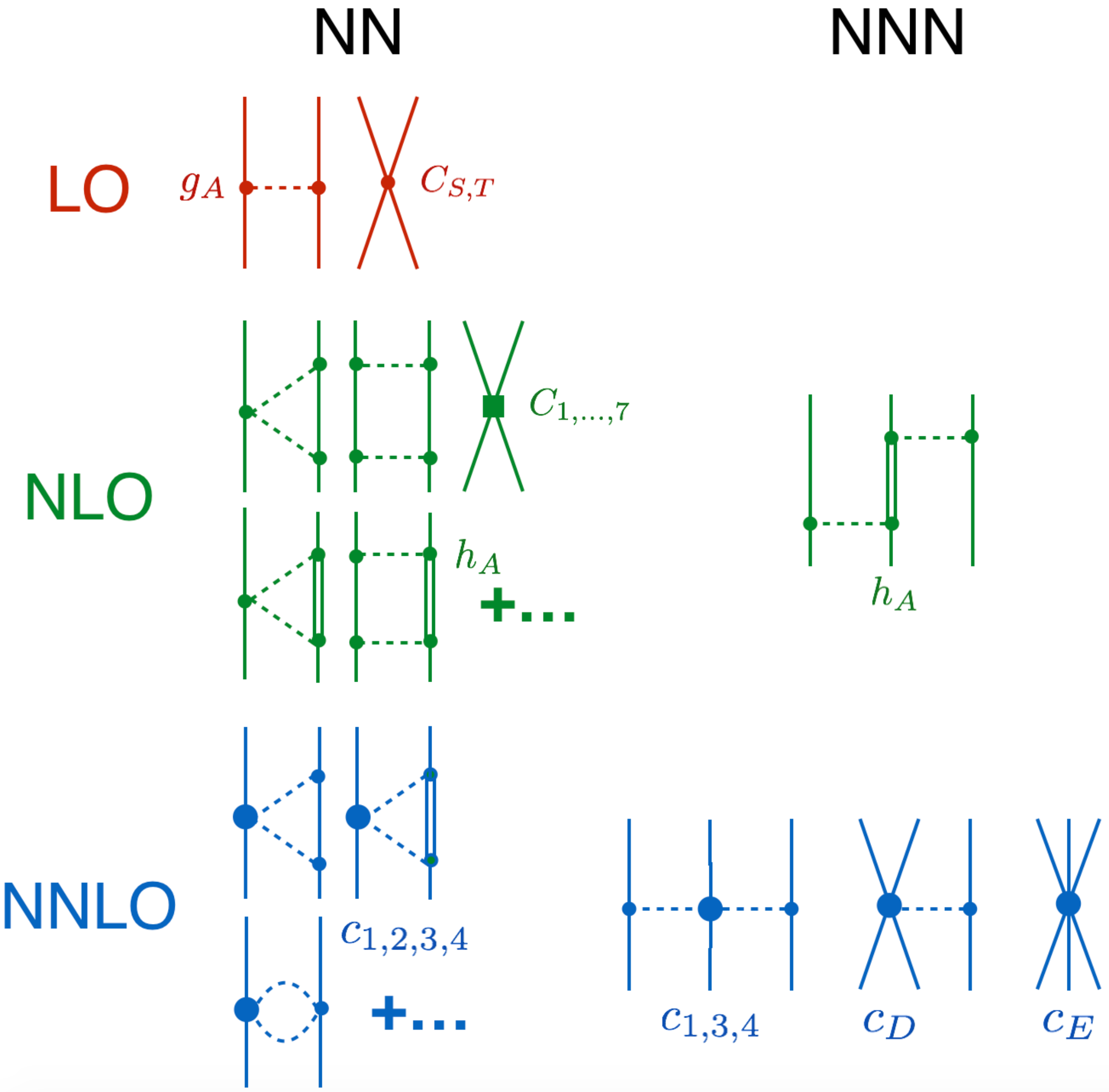}
\caption{(Color online) Schematic figure of relevant diagrams that enter in
$\Delta$-full \xEFT{} at leading order (LO), next-to-leading order
(NLO), and next-to-next-to-leading order (NNLO). The leading $\pi N$
and $\pi N \Delta$ axial couplings are denoted by $g_A$ and $h_{A}$,
respectively. Note that there are no $\Delta$ contributions at LO. At
NLO, the $NN$ contact interactions also remain unchanged. However, the
leading-order $NNN$ interaction, i.e. the well-known Fujita-Miyazawa
term~\cite{fujita1957}, appears at this order, see also
Ref.~\cite{epelbaum2008}. The $\Delta$ contributions to the $NNN$
interaction at NNLO vanish due to the Pauli principle or are
suppressed and demoted to a higher chiral order. Besides the
subleading $\pi N$ LECs $c_{1},c_{2},c_{3},c_{4}$, the $NNN$ diagrams
contain two additional LECs; $c_{D}$ and $c_{E}$.}
\label{delta_eft}
\end{center}
\end{figure}

Recently, \citeauthor{Piarulli2015} produced minimally non-local
\xEFT{} $NN$ potentials at next-to-next-to-next-to leading order
(N$^3$LO), with $\Delta$'s included up to next-to-next-to leading
order (NNLO), using values for the subleading $\pi N$ LECs
$c_1,c_2,c_3,c_4$ from Ref.~\cite{Krebs2007}. Dropping the non-local
terms led to the local potentials of Ref.~\cite{Piarulli2016}. Two
different approaches augmented these local potentials with $NNN$
forces up to NNLO. The corresponding diagrams of the $NNN$ force, some
$NN$ diagrams, and the most relevant LECs are shown in
Fig.~\ref{delta_eft}. \citeauthor{Logoteta2016} adjusted the LECs
$c_D$ and $c_E$ of the short-ranged $NNN$ terms to reproduce the
saturation point of nuclear matter.  However, they did not report
results for few-nucleon systems. In contrast,
\citeauthor{piarulli2017} adjusted $c_D$ and $c_E$ to reproduce
properties of nuclear systems with mass number $A=3$. Their quantum
Monte Carlo calculations yielded accurate results for spectra of light
nuclei up to $^{12}$C.  We note that the potentials by
\citeauthor{Logoteta2016} and \citeauthor{piarulli2017} employ values
for $c_D$ and $c_E$ that differ in signs and magnitudes.

In this paper we present a systematic construction and comparative
analysis of non-local $\Delta$-full and $\Delta$-less \xEFT{}
potentials at LO, NLO, and NNLO, and report results for light- and
medium-mass nuclei, and infinite nucleonic matter.  We constrain the
relevant short-ranged LECs using experimental data from nuclear
systems with mass numbers $A =2,4$ and use $\pi N$ LECs determined in
a recent high-precision analysis~\cite{Siemens2017} based on the
Roy-Steiner equations~\cite{hoferichter2016}. We do not include any
additional contact operators beyond NNLO in Weinberg power
counting. We find that the resulting $\Delta$-full potentials yield
accurate charge radii and much improved binding energies for medium
mass nuclei, and reproduce the saturation point of symmetric nuclear
matter within estimated EFT-truncation errors. Furthermore, estimates
of the EFT-truncation errors furnish a discussion of the improved
convergence rate of the $\Delta$-full \xEFT{} expansion compared to
the $\Delta$-less theory.

\section{Optimization of interactions}
To isolate the effects of the $\Delta$ isobar in the description of
the saturation properties of nucleonic matter we compare our results
with $\Delta$-less \xEFT{} potentials at LO, NLO, and NNLO. Other than
the inclusion of the $\Delta$-isobar, the $\Delta$-full and the
$\Delta$-less interactions are constructed following identical
optimization protocols. For the description of the interaction we
build on
work~\cite{vankolck1994,Hemmert1998,Kaiser1998,Krebs2007,epelbaum2008}
and treat the $\Delta$-$N$ mass difference $\delta \equiv M_{\Delta} -
M_{\rm N}$ as an additional small scale. A power-counting for this
approach is provided by the so-called small-scale
expansion~\cite{Hemmert1998}. This is identical to the conventional
heavy-baryon formulation of \xEFT{} which is already used for
including the nucleon mass-scale without any $\Delta$ isobars.  The
$\Delta$-less pion-exchanges in the $NN$ sector up to NNLO are given
in Ref.~\cite{entem2015}. The expressions for the $NN$ contact
potentials at LO and NLO are given in e.g. Ref.~\cite{machleidt2011},
and the $\Delta$ contributions to the leading and sub-leading
$2\pi$-exchanges in the $NN$ potential are from
Ref.~\cite{Krebs2007}. Charge-independence breaking terms are included
in the LO contact LECs as well as the one-pion exchange. Following
Ref.~\cite{Long2011} we remove all contributions that are proportional
to the subleading $\pi N \Delta$ coupling $b_3+b_8$ by renormalizing
the $\pi N \Delta$ axial coupling $h_A$ and the subleading $\pi N$
couplings $c_{2,3,4}$.  We follow \citeauthor{Siemens2017} and use
$h_{A} = 1.40$, $g_{A}=1.289$, and the central Roy-Steiner values of
the $\pi N$ LECs for the $\Delta$-full and $\Delta$-less potentials up
to third order. We recall that $\Delta$-less \xEFT~potentials often
employ $\pi N$ LECs with values that differ from what is found in $\pi
N$ scattering, because the absence of $\Delta$'s strongly renormalize
the $\pi N$ couplings $c_{2,3,4}$ in the three-nucleon
sector~\cite{pandharipande2005}. The $\Delta$-full theory is more
consistent in this regard and the $c_i$'s appear to be more natural in
size.

The expressions for the three-nucleon diagrams at NNLO are from
Ref.~\cite{epelbaum2002}. The NLO $NNN$-force in the $\Delta$-full
theory is given by the well-known Fujita-Miyazawa
term~\cite{fujita1957}. This topology is identical in structure to the
$\Delta$-less $2 \pi$-exchange $NNN$ interaction when using the
resonance-saturation values for the relevant $\pi N$ LECs
\begin{equation*}
  c_{3}^{\Delta} = - 2 c_{4}^{\Delta} = \frac{4h_{A}^2}{9\delta} = -2.972246\,\,{\rm GeV}^{-1}.
\end{equation*}

To construct quantitative $\Delta$-full \xEFT~potentials we need to
determine the numerical values of the LECs in the LO and NLO contact
potentials and the $c_D$ and $c_E$ terms in the $NNN$ interaction at
NNLO. To optimize the contact LECs we use a Levenberg-Marquardt
algorithm with machine-precise derivatives from automatic
differentiation~\cite{carlsson2016}. The objective function for the LO
and NLO contact LECs consists of the sum of squared differences
between the theoretical partial-wave $NN$ scattering phase shifts and
the corresponding values from the the Granada
analysis~\cite{Navarro2013} up to 200~MeV scattering energy in the
laboratory system. At LO, we only use phase shifts up to 1~MeV. The
neutron-neutron LEC $\tilde{C}_{^{1}S_{0}}^{(nn)}$ is constrained to
reproduce the effective range expansion in the $^1S_{0}$ channel. At
NNLO we use the same optimization algorithm to find the $c_D$ and
$c_E$ LECs that simultaneously reproduce the binding energy and
point-proton radius of $^{4}$He. Although correlated, these $A=4$
observables provide enough information to identify a unique minimum in
the $c_D$-$c_E$ plane that is sufficient for the purpose of comparing
the effects in nuclei and nucleonic matter due to the $\Delta$
isobar. An extended regression analysis or Bayesian inference approach
including additional data from many-nucleon systems or three-nucleon
scattering would generate interactions for use in detailed analyses of
atomic nuclei or model selection. In this work we focus on the effects
of the $\Delta$-isobar in nucleonic matter.

To regulate the the interactions we use the usual non-local regulators
\begin{align}
  \nonumber
  f(p)&={\rm exp}   \left[ -\left( \frac{p^2}{\Lambda^2}\right)^{3}       \right] \\
  \nonumber
  f(p,q)&= {\rm exp}\left[ -\left( \frac{4p^2+3q^2}{4\Lambda^2} \right)^3 \right]
\end{align}
in the $NN$ and $NNN$ interactions, resepectively. Here, $p$ and $q$
denote the Jacobi momenta in the two-body system and spectator
nucleon, respectively, and $\Lambda$ is the momentum cutoff. The
non-local regulator acts multiplicative, i.e.
\begin{eqnarray*}
  V_{NN}(p',p) &\to& f(p')V_{NN}(p',p)f(p) , \nonumber\\
  V_{NNN}(p',q';p,q) &\to& f(p',q')V_{NNN}(p',q';p,q)f(p,q) .
\end{eqnarray*}

\begin{table*}[htb]
  \caption{\label{tab:DLECs} Numerical values of the LECs for
    $\Delta$-full \xEFT{} potentials with a momentum cutoff
    $\Lambda=450$ MeV at LO, NLO, and NNLO. The $\pi N$ LECs
    $c_{1,2,3,4}$ are taken from the Roy-Steiner analysis in
    Ref.~\cite{Siemens2017}, and for consistency we use $h_{A}=1.40$,
    $g_{A}=1.289$, and $F_{\pi}=92.2$ MeV.}
  \begin{ruledtabular}
    \begin{tabular}{l|dddddd}
      LEC & \multicolumn{1}{c}{LO(450)} & \multicolumn{1}{c}{$\Delta$NLO(450)} & \multicolumn{1}{c}{$\Delta$NNLO(450)} & \multicolumn{1}{c}{LO(500)} & \multicolumn{1}{c}{$\Delta$NLO(500)} & \multicolumn{1}{c}{$\Delta$NNLO(500)}  \\ \hline
      $c_{1}$                     & - & - & -0.74 & - & - & -0.74 \\
      $c_{2}$                     & - & - & -0.49 & - & - & -0.49 \\
      $c_{3}$                     & - & - & -0.65 & - & - & -0.65 \\
      $c_{4}$                     & - & - & +0.96 & - & - & +0.96 \\ \hline
      $\tilde{C}_{^{1}S_{0}}^{(nn)}$ & -0.112927 & -0.310511 & -0.338023 & -0.108522 & -0.310256 & -0.338223 \\
      $\tilde{C}_{^{1}S_{0}}^{(np)}$ & -0.112927 & -0.310712 & -0.338139 & -0.108522 & -0.310443 & -0.338320 \\
      $\tilde{C}_{^{1}S_{0}}^{(pp)}$ & -0.112927 & -0.309893 & -0.337137 & -0.108522 & -0.309618 & -0.337303 \\
      $\tilde{C}_{^{3}S_{1}}$       & -0.087340 & -0.197951 & -0.229310 & -0.068444 & -0.191013 & -0.221721 \\ \hline
      $C_{^{1}S_{0}}$               & - & +2.391638 & +2.476589 & - & +2.395375 & +2.488019 \\
      $C_{^{3}S_{1}}$               & - & +0.558973 & +0.695953 & - & +0.539378 & +0.675353 \\
      $C_{^{1}P_{1}}$               & - & +0.004813 & -0.028541 & - & +0.015247 & -0.012651 \\
      $C_{^{3}P_{0}}$               & - & +0.686902 & +0.645550 & - & +0.727049 & +0.698454 \\
      $C_{^{3}P_{1}}$               & - & -1.000112 & -1.022359 & - & -0.951417 & -0.937264 \\
      $C_{^{3}P_{2}}$               & - & -0.808073 & -0.870203 & - & -0.793621 & -0.859526 \\
      $C_{^{3}S_{1}-^{3}D_{1}}$       & - & +0.362094 & +0.358330 & - & +0.358443 & +0.354479 \\ \hline
      $c_{D}$                     & - & - & +0.790  & - & - & -0.820 \\
      $c_{E}$                     & - & - & +0.017 & - & - & -0.350 \\
    \end{tabular}
  \end{ruledtabular}
\end{table*}

\begin{table*}[htb]
  \caption{\label{tab:LECs} Numerical values of the LECs for
    $\Delta$-less \xEFT{} potentials with a momentum cutoff
    $\Lambda=500$ MeV at LO, NLO, and NNLO. The $\pi N$ LECs
    $c_{1,3,4}$ are taken from the Roy-Steiner analysis in
    Ref.~\cite{Siemens2017}, and for consistency we use $g_{A}=1.289$,
    and $F_{\pi}=92.2$ MeV}
  \begin{ruledtabular}
    \begin{tabular}{l|dddddd}
      LEC & \multicolumn{1}{c}{LO(450)} & \multicolumn{1}{c}{NLO(450)} & \multicolumn{1}{c}{NNLO(450)} & \multicolumn{1}{c}{LO(500)} & \multicolumn{1}{c}{NLO(500)} & \multicolumn{1}{c}{NNLO(500)}  \\ \hline
      $c_{1}$                     & - & - & -0.74 & - & - & -0.74 \\
      $c_{3}$                     & - & - & -3.61 & - & - & -3.61 \\
      $c_{4}$                     & - & - & +2.44 & - & - & +2.44 \\ \hline
      $\tilde{C}_{^{1}S_{0}}^{(nn)}$ & -0.112927 & -0.149559 & -0.152421 & -0.108522 & -0.148625 & -0.152130 \\
      $\tilde{C}_{^{1}S_{0}}^{(np)}$ & -0.112927 & -0.150034 & -0.152630 & -0.108522 & -0.149167 & -0.152327 \\
      $\tilde{C}_{^{1}S_{0}}^{(pp)}$ & -0.112927 & -0.149336 & -0.151775 & -0.108522 & -0.148236 & -0.151463 \\
      $\tilde{C}_{^{3}S_{1}}$       & -0.087340 & -0.152884 & -0.166118 & -0.068444 & -0.147784 & -0.158592 \\ \hline
      $C_{^{1}S_{0}}$               & - & +1.438619 & +2.391093 & - & +1.479889 & +2.394670 \\
      $C_{^{3}S_{1}}$               & - & -0.684095 & +0.446631 & - & -0.692660 & +0.426020 \\
      $C_{^{1}P_{1}}$               & - & +0.305070 & +0.150981 & - & +0.304204 & +0.160280 \\
      $C_{^{3}P_{0}}$               & - & +1.207031 & +0.909408 & - & +1.225764 & +0.949224 \\
      $C_{^{3}P_{1}}$               & - & -0.386920 & -0.967768 & - & -0.385154 & -0.923166 \\
      $C_{^{3}P_{2}}$               & - & -0.167769 & -0.696173 & - & -0.137914 & -0.681166 \\
      $C_{^{3}S_{1}-^{3}D_{1}}$       & - & +0.132948 & +0.372585 & - & +0.133834 & +0.368968  \\ \hline
      $c_{D}$                     & - & - & +1.790 & - & - & +0.400 \\
      $c_{E}$                     & - & - & +0.130 & - & - & -0.270 \\
    \end{tabular}
  \end{ruledtabular}
\end{table*}

To explore the sensitivity of the results with respect to changes in
the cutoff $\Lambda$ we employ two common choices, namely
$\Lambda=450$~MeV and $\Lambda=500$~MeV. To regularize the $2
\pi$-exchanges in conjunction with non-local regulation we use the
standard spectral-function regularization (SFR)~\cite{epelbaum2005}
with a cutoff $\tilde{\Lambda}=700$ MeV throughout. It should also be
pointed out that recent work,
e.g. Refs.~\cite{valderrama2009,valderrama2011,epelbaum2015b},
indicates that a carefully selected local regulation of the
long-ranged $2\pi$-exchanges render SFR redundant and yields an
improved analytical structure of the scattering amplitude. However,
the overall existence of such scheme dependencies~\cite{dyhaldo2016} will
persist as long as the chiral interactions cannot be order-by-order
renormalized, see e.g Ref.~\cite{hoppe2017} for a recent analysis. The
numerical values of the employed $\pi N$ LECs and the optimized
short-ranged LECs for the $\Delta$-less as well as the $\Delta$-full
potentials are given in Tables~\ref{tab:DLECs} and \ref{tab:LECs}. For
the masses of the pions $(\pi^{\pm,0})$, proton, neutron, nucleon
($p,n,N$), and $\Delta$ we use the following values (in MeV):
$M_{\pi^{\pm}}=139.57018$, $M_{\pi^{0}}=134.9766$, $M_{p}=938.272046$,
$M_{n}=939.565379$, $M_{N}=938.918267$, and $M_{\Delta}=1232$,
respectively.

The statistical error from the Roy-Steiner analysis of the $\pi N$
scattering data, documented in Ref.~\cite{Siemens2017}, as well as
uncertainties due to the fit of the contact potentials, are not
considered any further in this work. When contrasted with the much
larger systematic uncertainties due to the truncation of the EFT, such
statistical errors presently play a lesser
role~\cite{carlsson2016,ekstrom2015b,Navarro2015}. It is important to
note that although the $\pi N$ LECs are extracted from $\pi N$ data
using a high-precision Roy-Steiner analysis, the corresponding LECs in
the $\Delta$-full sector are less precise due to the large uncertainty
in the underlying determination of $h_{A}$.

To provide a crude estimate of the EFT-truncation uncertainty we follow
Refs.~\cite{epelbaum2015,furnstahl2015} and write the EFT expansion
for an observable $X$ as $X = X_{0}\sum_{n=0}^{\infty} a_n
Q^{n}$. Here $X_{0}$ is the scale of the observable, given e.g. by the
LO prediction, $a_n$ are dimensionless expansion coefficients (with
$a_1=0$ in Weinberg power counting), and $Q\equiv p/\Lambda_{b}$ is
the ratio of the typical momentum $p$ and the breakdown momentum
$\Lambda_b$. The application of Bayes theorem with
boundless and uniform prior distribution of the expansion coefficients
$a_n$ leads to an expression for the truncation error at order NjLO
($j=0$: LO, $j=1$: NLO, $j=2$: NNLO) according to
\begin{equation}
\sigma_{X}({\rm NjLO}) = X_{0}Q^{j+2} {\rm max} (|a_{0}|,|a_{1}|,...,|a_{j+1}|) , 
\label{eq:eft_error}
\end{equation}
see Eq.~(36) of Ref.~\cite{furnstahl2015}. This estimate is in
semi-quantitative agreement with a Bayesian uncertainty quantification
of the truncation error. The uncertainty at LO is further constrained
to at least the size of the contribution of the higher chiral
orders. For the breakdown scale $\Lambda_{b}$, we start from
Ref.~\cite{epelbaum2015} but use a more conservative estimate of
$\Lambda_{b} = 500$~MeV. We also estimate the typical momentum-scale
for bound state observables as $p \sim m_{\pi}$, and employ $p \sim
p_{F}$ (the Fermi momentum) for infinite nucleonic matter, whereas for
$NN$ scattering we extract the momentum scale ${\rm max}(p_{\rm
  rel},m_{\pi})/\Lambda$. We disregard detailed numerical factors in
the various possible definitions of the relevant momentum scales for
bound states since the estimate in Eq.~(\ref{eq:eft_error}) is only
valid up to factors of order unity.

\begin{figure*}[htb]
  \includegraphics[width=0.7\textwidth]{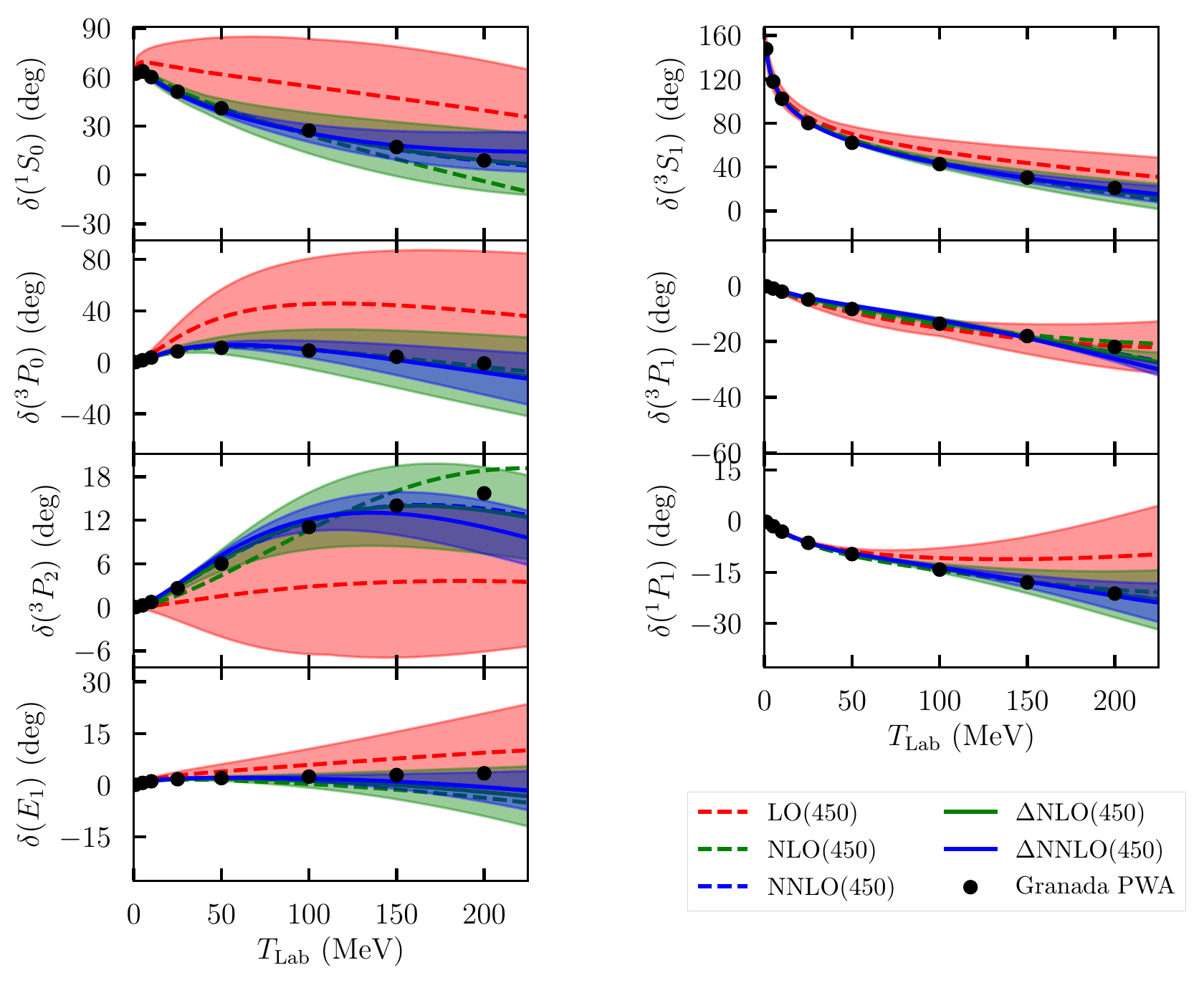}
  \caption{\label{fig:phases_np_cont_450}(Color online) Neutron-proton
    scattering phase shifts for the contact partial waves using the
    $\Delta$-full and $\Delta$-less \xEFT{} potentials with non-local cutoff
    $\Lambda=450$ MeV. All phases are compared to the results from the
    Granada phase shift analysis~\cite{Navarro2013}. The bands
    correspond to the order-by-order EFT truncation error in the
    $\Delta$-full approach, as described in the main text.}
\end{figure*}

\begin{figure*}[htb]
  \includegraphics[width=0.7\textwidth]{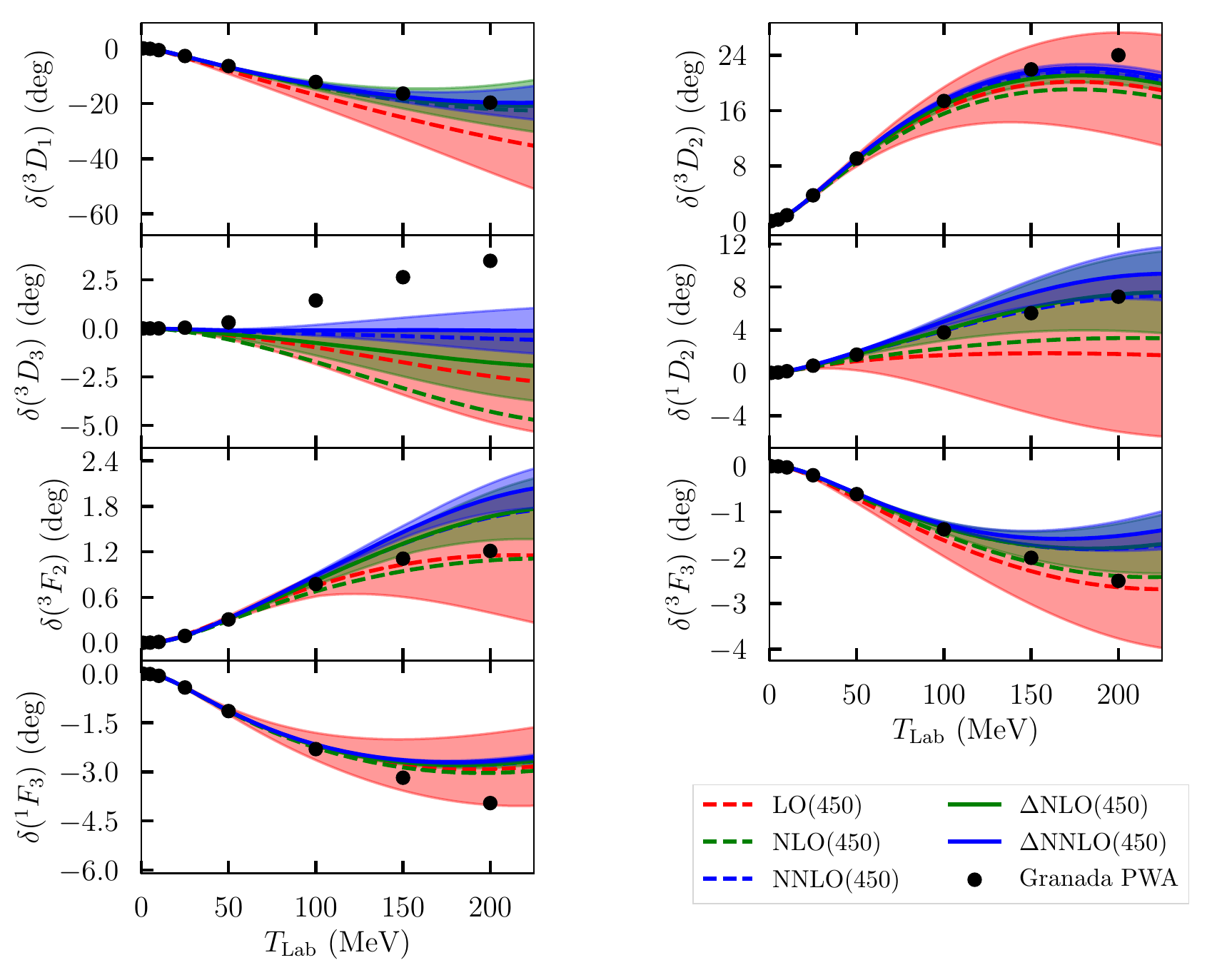}
  \caption{\label{fig:phases_np_periph_450}(Color online) Neutron-proton
    scattering phase shifts for selected peripheral partial waves using the
    $\Delta$-full and $\Delta$-less \xEFT{} potentials with non-local cutoff
    $\Lambda=450$ MeV. All phases are compared to the results from the
    Granada phase shift analysis~\cite{Navarro2013}. The bands
    correspond to the order-by-order EFT truncation error in the
    $\Delta$-full approach, as described in the main text.}
\end{figure*}

\begin{figure*}[htb]
  \includegraphics[width=0.7\textwidth]{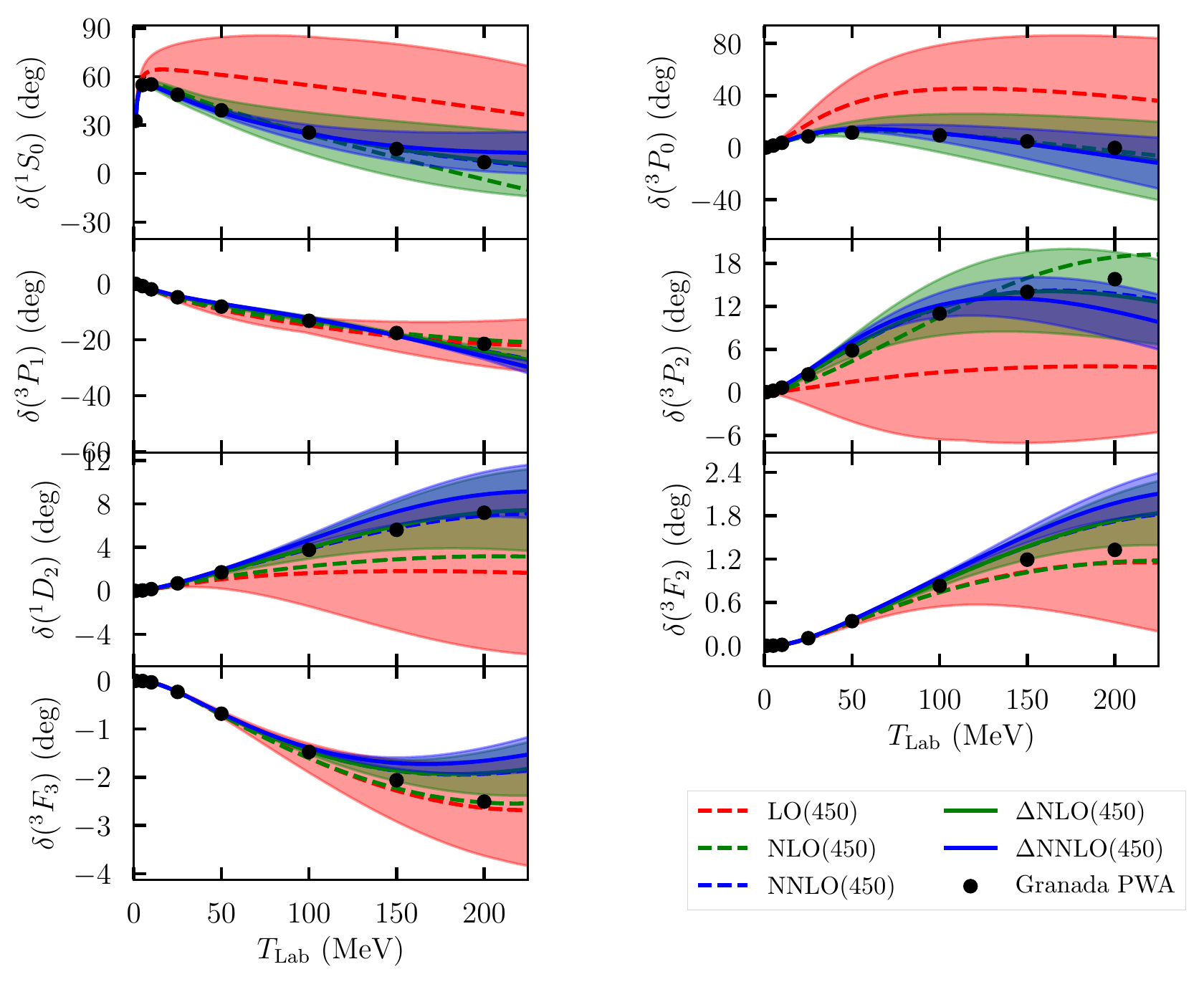}
  \caption{\label{fig:phases_pp_all_450}(Color online) Proton-proton
    scattering phase shifts for the contact and selected peripheral
    partial waves using the $\Delta$-full and $\Delta$-less \xEFT{}
    potentials with non-local cutoff $\Lambda=450$ MeV. All phases are
    compared to the results from the Granada phase shift
    analysis~\cite{Navarro2013}. The bands correspond to the
    order-by-order EFT truncation error in the $\Delta$-full approach,
    as described in the main text.}
\end{figure*}

In Figs.~\ref{fig:phases_np_cont_450},~\ref{fig:phases_np_periph_450},
and ~\ref{fig:phases_pp_all_450} we compare the quality of the $NN$
scattering phase shifts of the $\Delta$-full and $\Delta$-less
interactions with cutoff $\Lambda=450$ MeV. The results for the
peripheral waves agree well with published interactions that were
analyzed in the Born approximation~\cite{Krebs2007}. The dashed lines
show the $\Delta$-less results, order by order from red to green to
blue.  The full lines show the $\Delta$-full results, and we remind
the reader that LO is not affected by the $\Delta$ (see
Fig.~\ref{delta_eft}).
Clearly, in several partial waves the $\Delta$-full \xEFT~interactions
exhibit a faster order-by-order convergence than the corresponding
$\Delta$-less formulations.  Somewhat surprisingly, the \DNNLO{}
results at higher scattering energies, in particular for $^{1}$S$_{0}$
and selected peripheral waves, such as $^{1}$D$_{2}$, are slightly
less accurate than the corresponding $\Delta$-less order. A more
involved optimization strategy, such as Bayesian parameter estimation,
could further illuminate this point. Nevertheless, the Granada phase
shifts fall on the envelope of the estimated truncation errors and the
results therefore seem reasonable. Although not shown,
the computed phase shifts for the $\Lambda=500$ MeV interactions are
very similar and exhibit the same features.

Tables~\ref{tab:ncsm450} and ~\ref{tab:ncsm500} summarize our
results for selected bound-state observables in $A \leq 4$ nuclei
computed with a Jacobi-coordinate version~\cite{kamuntavicius2000}
of the no-core shell model (NCSM)~\cite{navratil2009,barrett2013}. All
calculations are converged in 41 and 21 major oscillator shells with
$\hbar \Omega=36$~MeV for $A=3$ and $A=4$, respectively. The charge
radius and binding energy of $^4$He were used to constrain the LECs
$c_D$ and $c_E$ of the short-ranged three-nucleon force whereas the
NCSM results for $A=2,3$ nuclei are predictions. At NNLO, all results
except the binding energy of $^{2}$H, agree with the experimental
values within the estimated EFT-truncation errors. The computed
point-proton radii were transformed to charge radii using a standard
expression, see e.g. Ref.~\cite{ekstrom2015a}.

\begin{table*}[hbt]
  \caption{\label{tab:ncsm450} Binding energies ($E$) in MeV, charge
    radii ($R_{\rm ch}$) in fm, for $^{2,3}$H and $^{3,4}$He at LO,
    NLO, and NNLO~with $\Lambda=450$ MeV, with and without the
    $\Delta$-isobar and compared to experiment. For the ground state
    of $^{2}$H we also present the quadrupole moment ($Q$) in e
    fm$^{2}$ and the $D$-state probability ($P_{\rm D}$) in
    \%. Experimental charge radii are from
    Ref.~\cite{angeli2013}. Estimates of the EFT truncation-errors are
    given in parenthesis, and at LO we report the truncation error
    belonging to the $\Delta$-full expansion.}
\begin{ruledtabular}
\begin{tabular}{l|D{.}{.}{2.7}D{.}{.}{2.7}D{.}{.}{2.7}D{.}{.}{2.7}D{.}{.}{2.7}D{.}{.}{2.7}D{.}{.}{2.7}}
& \multicolumn{1}{c}{LO(450)} & \multicolumn{1}{c}{NLO(450)} & \multicolumn{1}{c}{$\Delta$NLO(450)} & \multicolumn{1}{c}{NNLO(450)} & \multicolumn{1}{c}{$\Delta$NNLO(450)} & \multicolumn{1}{c}{Exp.} \\ \hline
$E$($^2$H) & 2.01(15) & 2.02(12) & 2.10(5) & 2.14(3) & 2.16(2) & 2.2245\\
$R_{\rm ch}$($^2$H)  &2.16(16) & 2.167(16) & 2.156(7) & 2.1511(44) & 2.1486(19) & 2.1421(88)\\
$P_{\rm D}$($^2$H)  & 7.15(3.51) & 3.43(1.02) & 3.63(97) & 3.70(28) &  3.74(27) &  - \\
$Q$($^2$H)          & 0.322(41)  &  0.276(13) & 0.277(11)  & 0.277(3) &  0.277(3)  & 0.27\footnote{CD-Bonn value~\cite{machleidt2011}}\\ \hline
$E$($^3$H) & 10.91(2.38) & 8.54(65) &  8.65(62) &  8.56(18) & 8.53(17) & 8.48\\
$R_{\rm ch }$($^3$H)& 1.52(23) &  1.70(5) & 1.72(6) &  1.74(1) & 1.74(2) & 1.7591(363)\\ \hline
$E$($^3$He) & 9.95(2.21) &  7.78(60) & 7.85(58) &  7.78(16) & 7.73(16) & 7.72 \\
$R_{\rm ch}$($^3$He) & 1.66(32) &  1.91(7) & 1.94(8) &  1.96(2) & 1.97(2) & 1.9661(30)\\ \hline
$E$($^4$He) & 39.60(11.3) &  30.10(2.62) & 29.32(2.83) &  28.30(72) & 28.29(78) & 28.30\\
$R_{\rm ch }$($^4$He) & 1.37(30) &  1.59(9) & 1.63(7) &  1.68(3) & 1.67(2) & 1.6755(28)\\
\end{tabular}
\end{ruledtabular} 
\end{table*}

\begin{table*}[htb]
  \caption{\label{tab:ncsm500} Binding energies ($E$) in MeV, charge
    radii ($R_{\rm ch}$) in fm, for $^{2,3}$H and $^{3,4}$He at LO,
    NLO, and NNLO~with $\Lambda=500$~MeV, with and without the
    $\Delta$-isobar and compared to experiment. For the ground state
    of $^{2}$H we also present the quadrupole moment ($Q$) in e
    fm$^{2}$ and the $D$-state probability ($P_{\rm D}$) in
    \%. Experimental charge radii are from
    Ref.~\cite{angeli2013}. Estimates of the EFT truncation-errors
    given in the parenthesis, and at LO we report the truncation error
    belonging to the $\Delta$-full expansion.}
  \begin{ruledtabular}
    \begin{tabular}{l|D{.}{.}{2.7}D{.}{.}{2.7}D{.}{.}{2.7}D{.}{.}{2.7}D{.}{.}{2.7}D{.}{.}{2.7}}
      & \multicolumn{1}{c}{LO(500)} & \multicolumn{1}{c}{NLO(500)} & \multicolumn{1}{c}{$\Delta$NLO(500)} & \multicolumn{1}{c}{NNLO(500)} & \multicolumn{1}{c}{$\Delta$NNLO(500)} & \multicolumn{1}{c}{Exp.} \\ \hline
      $E$($^2$H)            &2.04(16)  &  2.04(12)  & 2.12(5)  &  2.16(3)  & 2.18(2)   & 2.2245\\
      $R_{\rm ch}$($^2$H) & 2.15(16)    &  2.164(16)  & 2.153(7) &  2.149(4)   & 2.1459(19) & 2.1421(88)\\
      $P_{\rm D}$($^2$H)  & 7.80(3.97)  &  3.55(1.17)  & 3.82(1.09) &  3.93(32)    & 3.97(30)   &  - \\
      $Q$($^2$H)           & 0.317(42) &  0.273(12)   & 0.276(11) &  0.275(3)  & 0.276(3)   & 0.27\footnote{CD-Bonn value~\cite{machleidt2011}}\\ \hline
      $E$($^3$H)             & 10.47(1.97) &  8.42(56) & 8.91(43) &  8.49(16)  & 8.50(12)  & 8.48\\
      $R_{\rm ch }$($^3$H) & 1.54(21)    &  1.71(5)  & 1.71(5)    &  1.75(1)  & 1.75(1)    & 1.7591(363)\\ \hline
      $E$($^3$He)            & 9.50(1.80) &  7.66(51) & 8.11(40)  &  7.72(14)  & 7.70(11)  & 7.72 \\
      $R_{\rm ch}$($^3$He) & 1.68(30) &  1.93(7)   & 1.92(7)    &  1.97(2)  & 1.98(2)    & 1.9661(30)\\ \hline
      $E$($^4$He)            & 37.00(8.69) &  29.22(2.15) & 30.70(2.38) &  28.31(60) & 28.31(65) & 28.30\\
      $R_{\rm ch }$($^4$He) & 1.39(28)  &  1.60(7)   & 1.62(6)    &  1.68(2)  & 1.67(2)    & 1.6755(28)\\
    \end{tabular}
  \end{ruledtabular}
  \end{table*}

\section{Predictions for medium mass nuclei and nucleonic matter}
In this Section we present results for selected finite nuclei and
infinite nucleonic matter. For nucleonic matter we present results for
both $\Delta$-less and $\Delta$-full interactions, while for finite
nuclei we limit the discussion to the $\Delta$-full interactions since
the $\Delta$-less interactions produce nuclei that are not bound with
respect to breakup into $\alpha$-particles. The computed binding
energies and radii of finite nuclei are consistent with our results
for the saturation point in symmetric nuclear matter.

\subsection{Finite nuclei}
The many-body calculations for finite nuclei are performed with the
coupled-cluster (CC)
method~\cite{kuemmel1978,bishop1991,bartlett2007,hagen2014}.
We employ
the translationally invariant Hamiltonian
\beq
\label{ham}
H= T-T_{\rm cm} +V_{NN}+V_{NNN} .
\eeq
Here, $T$ denotes the total kinetic energy and $T_{\rm cm}$ the
kinetic energy of the center of mass. As the Hamiltonian~(\ref{ham})
does not reference the center-of-mass coordinate, the ground-state
wave function is a product of an intrinsic and a Gaussian
center-of-mass wave
function~\cite{hagen2009a,hagen2010b,jansen2012,morris2015,hergert2016}.
The CC method yields a similarity transformed Hamiltonian whose ground
state is the product state corresponding to a closed-shell nucleus. In
the coupled-cluster singles and doubles (CCSD) approximation,
typically accounting for about 90\% of the correlation energy, the
ground-state is orthogonal to all 1-particle--1-hole ($1p$-$1h$) and
$2p$-$2h$ excitations. In addition to the CCSD approximation we
include leading-order $3p$-$3h$ excitations perturbatively by
employing the $\Lambda$-CCSD(T)
method~\cite{taube2008,hagen2010b,binder2013}. This approximation
typically captures about 99\% of the correlation energy. We employ a
model space of 15 oscillator shells with $\hbar\Omega=16$~MeV, and a
cutoff $E_{\rm 3 max}=16\hbar\Omega$ for the maximum excitation energy
of three nucleons interacting via the three-nucleon potential
$V_{NNN}$. This potential enters the CC calculations in the
normal-ordered two-body approximation~\cite{hagen2007a,roth2012} in
the Hartree-Fock basis.

\begin{figure}[htb]
\includegraphics[width=0.5\textwidth]{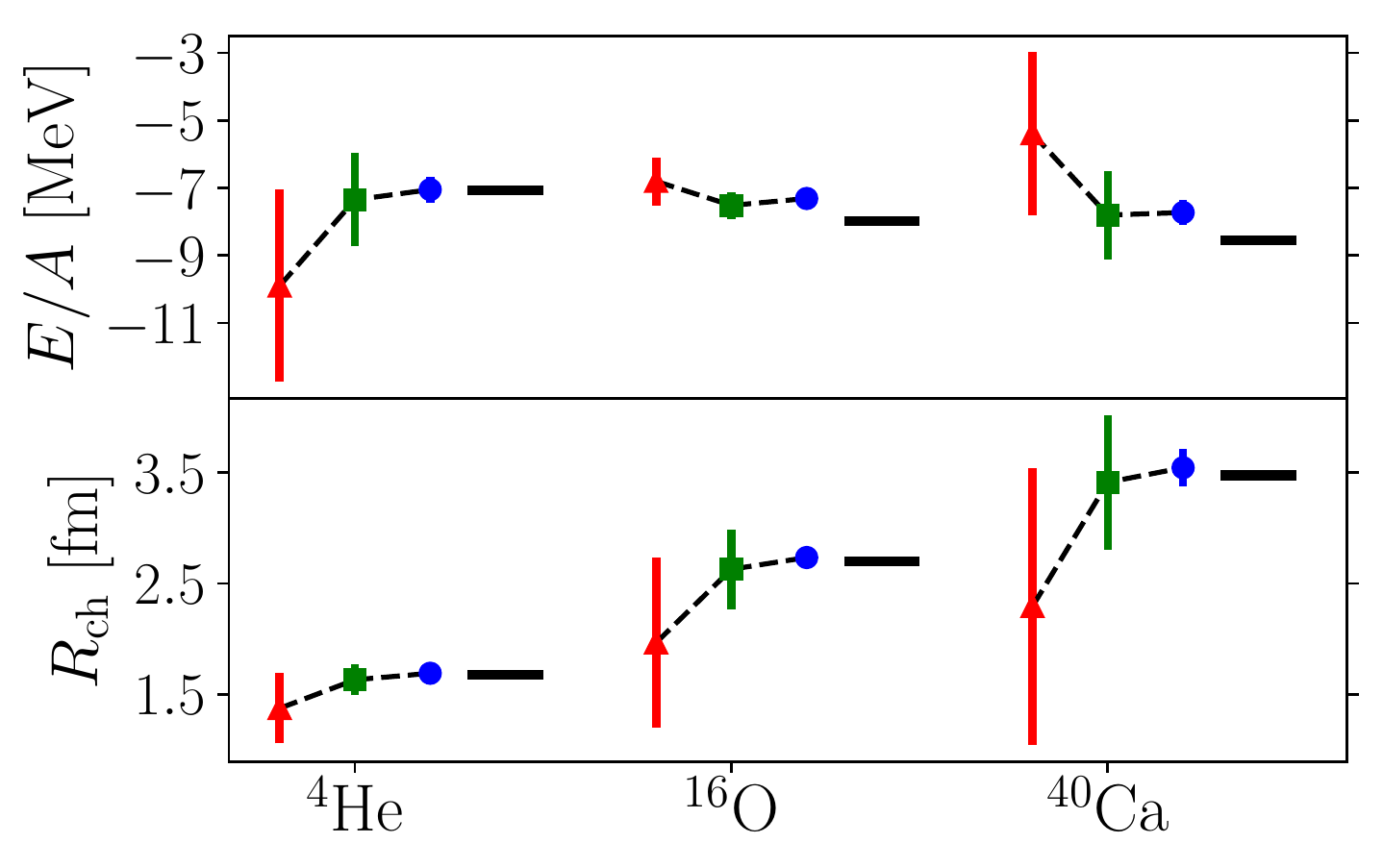}
\caption{(Color online) Ground-state energy (negative of binding
  energy) per nucleon and charge radii for selected nuclei computed
  with coupled cluster theory and the $\Delta$-full potential
  ($\Lambda=450$~MeV). For each nucleus, from left to right: LO (red
  triangle), NLO (green square), and NNLO (blue circle). The black
  bars are data. Vertical bars estimate uncertainties from the
  order-by-order EFT truncation errors $\sigma$(LO), $\sigma$(NLO),
  and $\sigma$(NNLO). At NLO and NNLO we estimate a conservative 95\%
  confidence interval, i.e $1.96 \times \sigma$. See the text for
  details.}
\label{fig:nuclei}
\end{figure}

To asses the impact of the $\Delta$-isobar in finite nuclei we
calculated the binding energies and charge radii for $^{4}$He,
$^{16}$O, and $^{40}$Ca order-by-order, i.e. at LO, NLO, and
NNLO. Figure~\ref{fig:nuclei} shows the results using the
$\Delta$-full interactions with a momentum cutoff $\Lambda=450$ MeV.
The ground-state energies are $E(^{16}{\rm O})=-108.8(11.4)$ , $-120.3
(6.4)$, and $-117.0(1.8)$~MeV and $E(^{40}{\rm Ca}) = -216 (97)$,
$-312 (52)$, and $-309 (14)$~MeV at LO, NLO, and NNLO, respectively.
The charge radii are $R_{\rm ch}(^{16}{\rm O}) = 1.96 (0.76)$, $2.63
(0.36)$, and $2.73 (0.10)$~fm and $R_{\rm ch}(^{40}{\rm Ca}) = 2.29
(1.25)$, $3.41 (0.61)$, and $3.55 (0.17)$~fm at LO, NLO, and NNLO,
respectively. Before we analyze the results, we estimate the
systematic uncertainties due to the truncation of the EFT. Again we
follow Refs.~\cite{epelbaum2015,binder2015}, use
Eq.~(\ref{eq:eft_error}) and set the momentum scale $p=m_{\pi}$ for
our low-energy observables. The predicted charge radii are accurate at
each order within uncertainties. Already at NLO, which is independent
of the sub-leading $2 \pi$-exchange LECs $c_{i}$, we obtain an
accurate description of both radii and binding energies of $^{4}$He,
$^{16}$O and $^{40}$Ca.  At NNLO, the charge radii also exhibit a
first sign of convergence in terms of the chiral expansion. Binding
energies exhibit a nearly identical order-by-order increase in
precision but somewhat underbind nuclei at NNLO. These results
demonstrate that the $\Delta$-isobar can play an important role also
in low-energy nuclear structure and nuclear
saturation~\cite{hebeler2011,ekstrom2015b}.

The $\Delta$ degree of freedom also impacts the stability of nclei
with respect to breakup into alpha particles. At LO, $^{16}$O and
$^{40}$Ca are not stable with respect to alpha emission. Similar
results were observed in pionless
EFT~\cite{stetcu2007,contessi2017,bansal2017} and nuclear lattice
EFT~\cite{elhatisari2016}.  However, the $\Delta$ modifies the
$2\pi$-exchanges between nucleons, and we observe that the
$\Delta$-full interactions at NLO and NNLO yield nuclei that are
stable with respect to alpha emission. This is in stark contrast to
results we obtained here using the $\Delta$-less NLO and NNLO
interactions at cutoff $\Lambda = 450$~MeV, and to those of
Ref.~\cite{carlsson2016}.

Table~\ref{tab_data} summarizes binding energies, radii, and also the
neutron skins of nuclei with closed subshells up to $^{48}$Ca. Note
that the lack of a spin-orbit (LS) force at LO results in
energy-degeneracies that hamper CC calculations of non LS-closed
nuclei. Therefore, we can obtain EFT truncation-errors only for
$^{16}$O and $^{40}$Ca using Eq.~(\ref{eq:eft_error}). For $^{48}$Ca
we predict a neutron skin of $R_{\rm skin} = 0.15$~fm at $\Delta$NLO
and $\Delta$NNLO, consistent with the recent ranges $0.14$-$0.20$~fm and
$0.12$-$0.15$~fm from Ref.~\cite{birkhan2017} and Ref.~\cite{hagen2015},
respectively.
\begin{table*}[t]
  \caption{\label{tab_data} Binding energies (E) (in MeV), charge radii (in fm), proton
   point radii (in
   fm), neutron point radii (in fm), and neutron skin (in fm) for $^{8}$He,
   $^{16,22,24}$O, and $^{40,48}$Ca at $\Delta$NLO and $\Delta$NNLO,
   and compared to experiment.}
\begin{ruledtabular}
\begin{tabular}{cddd|ddd|dd|dd|dd}
& \multicolumn{3}{c|}{$E$}    &\multicolumn{3}{c|}{$R_{\rm ch}$}&\multicolumn{2}{c}{$R_{\rm p}$} 
&\multicolumn{2}{c}{$R_{\rm n}$} &\multicolumn{2}{c}{$R_{\rm skin}$} \\ \hline
& \multicolumn{1}{c}{$\Delta$NLO}&\multicolumn{1}{c}{$\Delta$NNLO}&\multicolumn{1}{c|}{Exp.~\cite{wang2012}}
& \multicolumn{1}{c}{$\Delta$NLO}&\multicolumn{1}{c}{$\Delta$NNLO}&\multicolumn{1}{c|}{Exp. \cite{angeli2013}}
& \multicolumn{1}{c}{$\Delta$NLO}&\multicolumn{1}{c}{$\Delta$NNLO} 
& \multicolumn{1}{c}{$\Delta$NLO}&\multicolumn{1}{c}{$\Delta$NNLO}
&\multicolumn{1}{c}{$\Delta$NLO}&\multicolumn{1}{c}{$\Delta$NNLO} \\ \hline
$^{8}$He  &  27.5  & 27.0 &   31.40    &  1.90 &  1.97 & 1.924(31) & 1.77 & 1.85 & 2.63 & 2.70 & 0.85 &  0.85  \\
$^{16}$O  & 120.3  & 117.0&  127.62    &  2.63 &  2.73 & 2.699(5)  & 2.49 & 2.61 & 2.47 & 2.58 &-0.02 & -0.03 \\
$^{22}$O  & 146.2  & 145.4&  162.04    &  2.66 &  2.77 &           & 2.54 & 2.66 & 2.88 & 3.00 & 0.34 &  0.34 \\
$^{24}$O  & 152.2  & 151.6&  168.96    &  2.70 &  2.81 &           & 2.59 & 2.71 & 3.11 & 3.22 & 0.52 &  0.51 \\
$^{40}$Ca & 312.2  & 309.1&  342.05    &  3.41 &  3.55 & 3.478(2)  & 3.31 & 3.45 & 3.26 & 3.40 &-0.05 & -0.05 \\ 
$^{48}$Ca & 373.4  & 373.8&  416.00    &  3.45 &  3.56 & 3.477(2)  & 3.36 & 3.47 & 3.51 & 3.62 & 0.15 &  0.15 \\
\end{tabular}
\end{ruledtabular} 
\end{table*}

\begin{figure}[htb]
\includegraphics[width=0.5\textwidth]{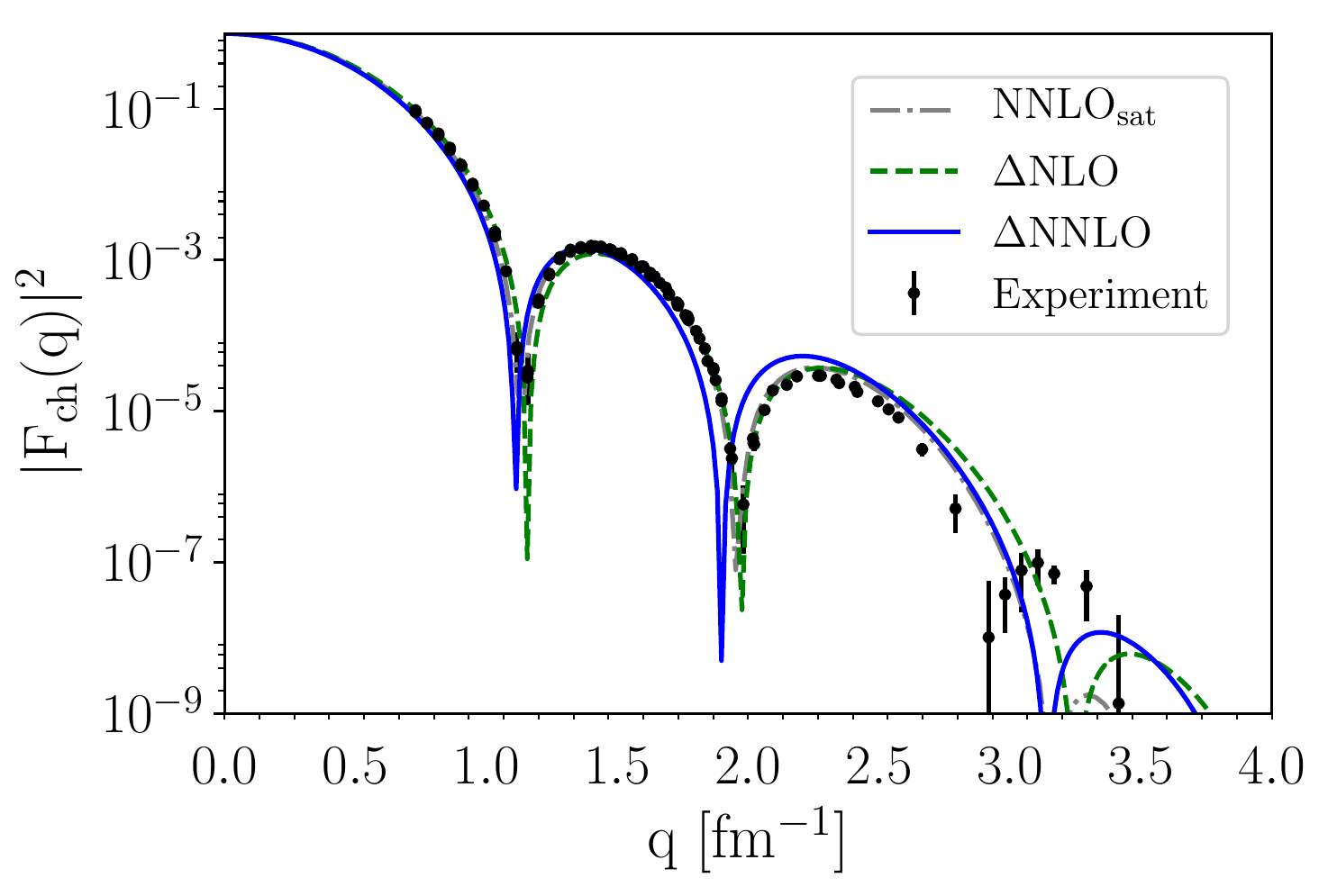}
\caption{(Color online) Elastic charge form factor of $^{48}$Ca from
  NNLO$_{\rm sat}$ (gray dash-dotted), $\Delta$NLO (green dashed), and
  $\Delta$NNLO (blue solid line) compared to experimental data
  (black).}
\label{fch2_48ca}
\end{figure}

Figure~\ref{fch2_48ca} shows the charge form-factor at $\Delta$NLO and
$\Delta$NNLO, compared to NNLO$_{\rm sat}$~\cite{ekstrom2015a} and
data. The charge form-factor is obtained by a Fourier transform of the
intrinsic charge density~\cite{giraud2008,hagen2015}, and agrees with
data for momentum transfers up to about $q\approx 2.5$~fm$^{-1}$. Also
for this quantity, the $\Delta$NLO results indicate an improved
convergence of chiral expansion compared to the $\Delta$-less
formulation.

We also computed spectra of various nuclei. These explorations
exhibited mixed results: While the low-lying states in $^{17}$O were
in good agreement with data, $^{25}$O is bound at $\Delta$NNLO with
respect to $^{24}$O by about~0.5MeV, and the $J^\pi=2^+$ state in
$^{24}$O is too low.  We believe that these shortcomings should not
distract from the main results reported in this work: accurate
saturation properties at NLO in the $\Delta$-full \xEFT. We speculate
that finer details such as spectra will require us to go to higher
order in the $NN$ interaction (as was done, e.g., in
Ref.~\cite{piarulli2017} by including N$^3$LO contacts), or to vary
the $\Delta$-full $\pi N$ couplings within their somewhat more
generous uncertainty limits due to the rather poorly known $\pi N
\Delta$ coupling $h_A$, or to also use data of heavier nuclei in the
optimization of the interaction. The interactions constructed in this
work serve as excellent starting points for such endeavors.

\subsection{Nucleonic matter}
We turn to the CC calculations of nuclear matter using $\Delta$-full
and $\Delta$-less interactions up to NNLO. We follow
Ref.~\cite{hagen2013b} and employ a Hamiltonian $H= T +V_{NN}
+V_{NNN}$.  The basis is a discrete lattice in momentum space
corresponding to periodic boundary conditions in a cubic box of length
$L$ in position space, and the discrete lattice momenta are given by
$2\pi \hbar n_i/L$, with $n_i = 0, \pm 1, \ldots \pm
n_{\mathrm{max}}$, and $i = x,y,z$.  We used $n_{\mathrm{max}} = 4$ as
the maximum number of lattice points. The CC calculations were carried
out at the doubles excitation level ($2p$-$2h$) with perturbative triples
($3p$-$3h$) corrections [CCD(T)]. Due to translational invariance,
there are no $1p$-$1h$ excitations. We use ``closed-shell'' lattice
configurations with 66 neutrons for neutron matter, and 132 nucleons
for symmetric nuclear matter. These nucleon numbers exhibit only small
finite-size effects~\cite{gandolfi2009,hagen2013b}. The CCD(T)
calculations were performed with the normal-ordered two-body
approximation for the $NNN$ interaction~\cite{hagen2007a,roth2012},
i.e. the three-nucleon force enters the normal-ordered Hamiltonian as
0-body, 1-body and 2-body interactions; summing over 3, 2, and 1
particles in the reference state, respectively. All results 
are well converged for $n_{\mathrm{max}} = 4$ at all considered
densities, i.e. $\rho\le 0.2$~fm$^{-3}$ To gauge the quality of the
normal-ordered two-body approximation, we also included the
``residual'' $NNN$ interaction (i.e. those that generate $3p$-$3h$
excitations when acting on the reference) in perturbation theory.  We
found that the residual $NNN$ contribution is negligible for neutron
matter, and small (0.2-0.3~MeV per nucleon) in symmetric nuclear
matter. This suggests that the normal-ordered two-body approximation
for the three-nucleon force is sufficiently precise for the
$\Delta$-full interactions considered in this work.
\begin{figure*}[htb]
  \includegraphics[width=0.45\textwidth]{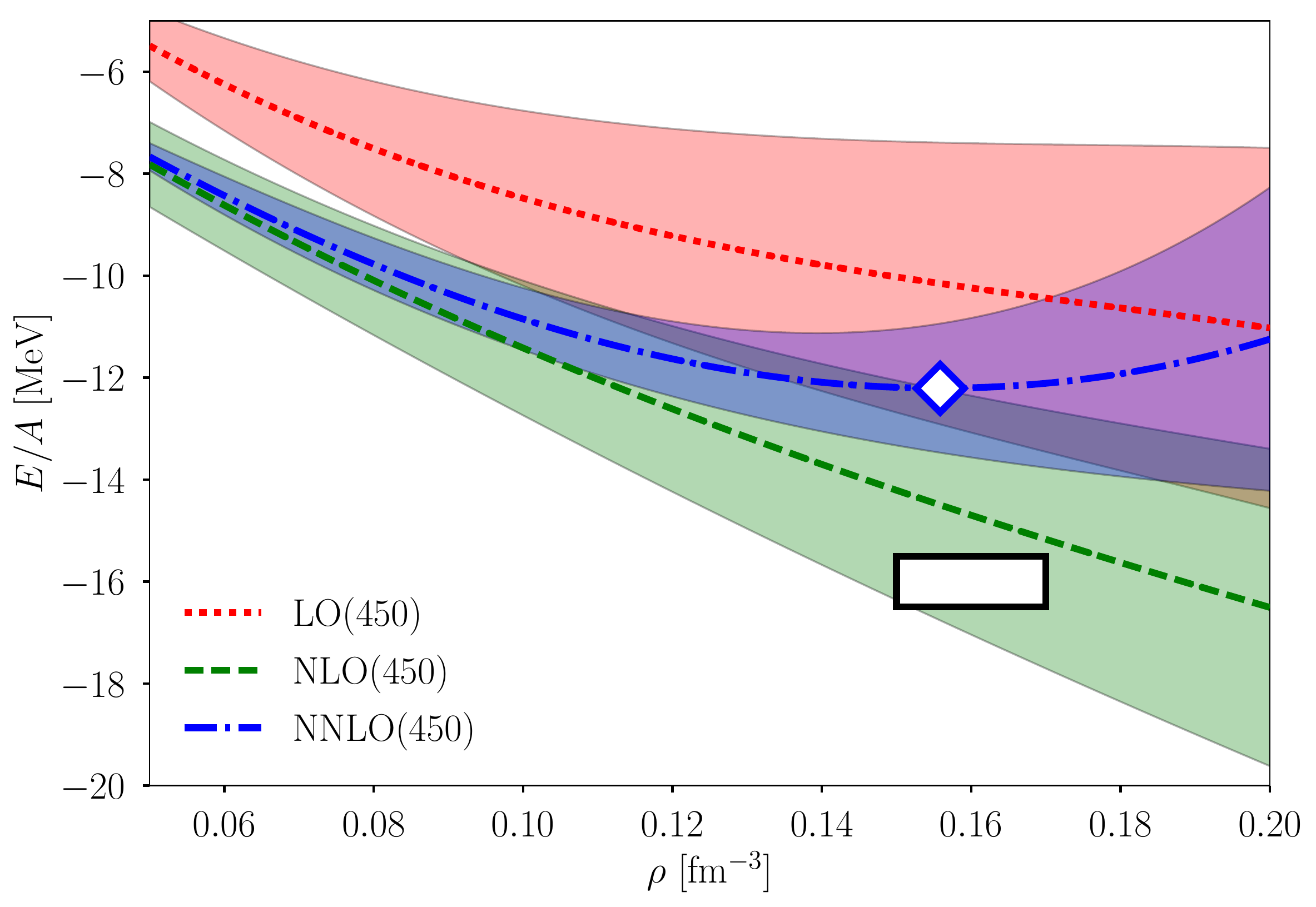}
  \includegraphics[width=0.45\textwidth]{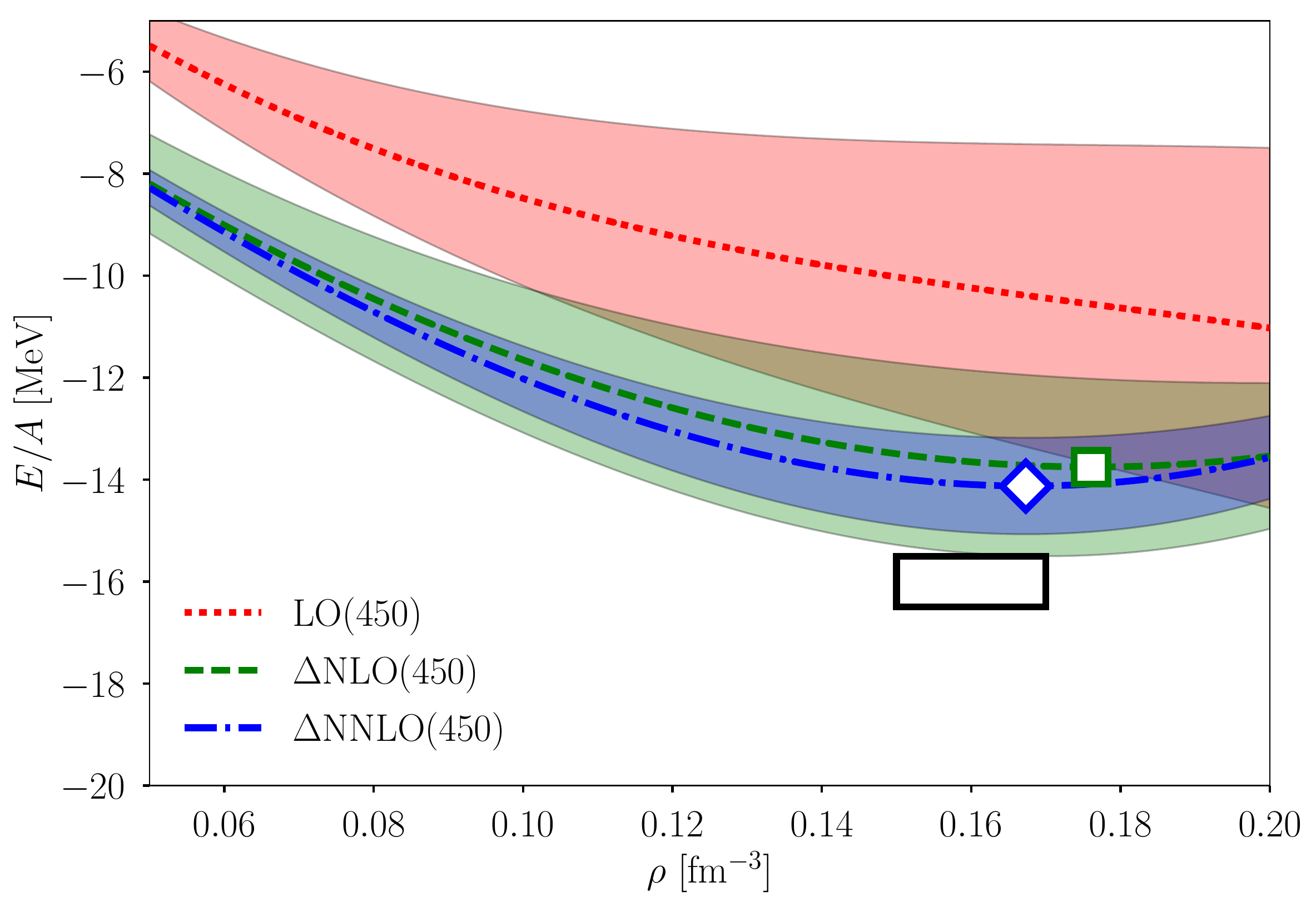}
  \includegraphics[width=0.45\textwidth]{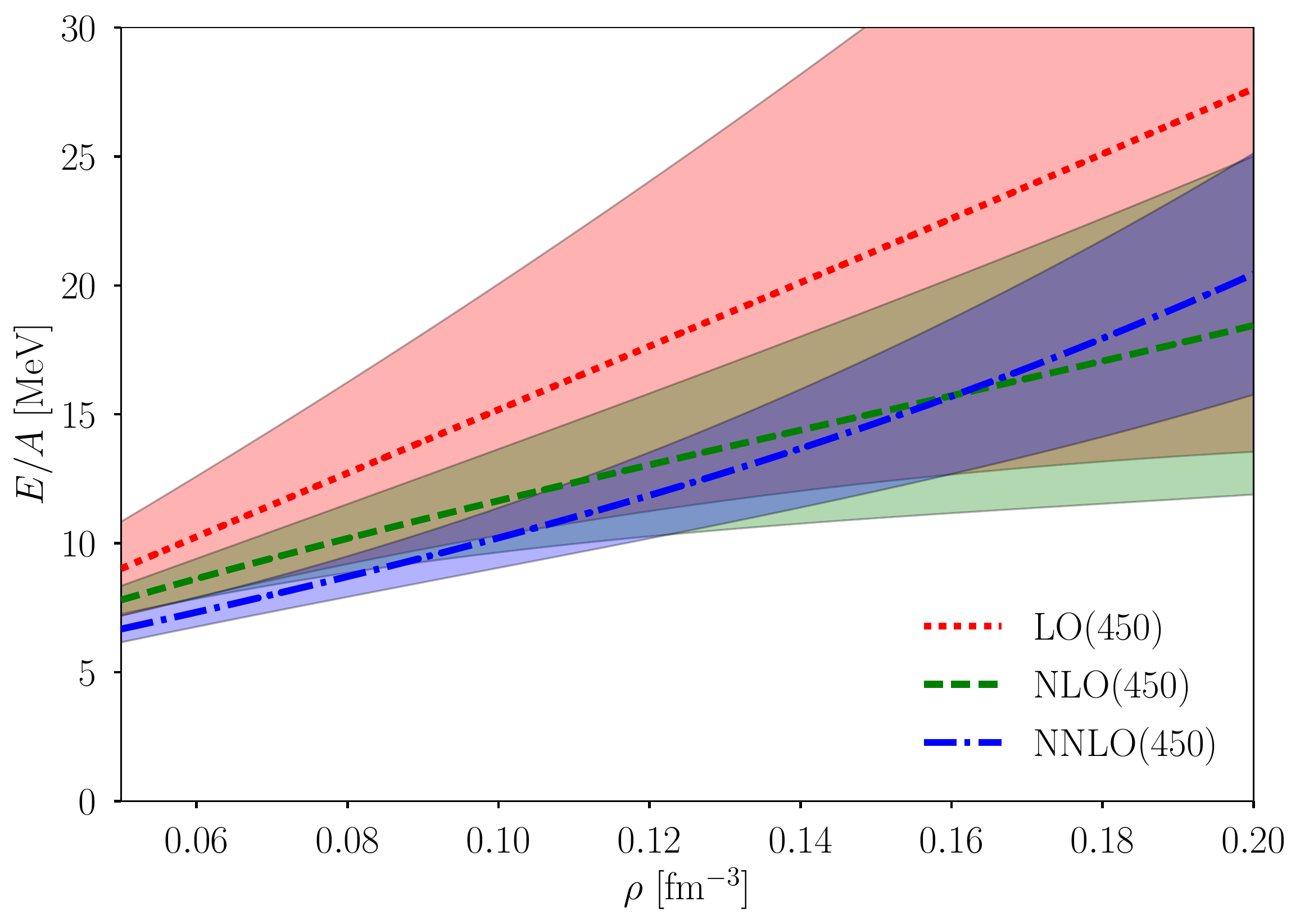}
  \includegraphics[width=0.45\textwidth]{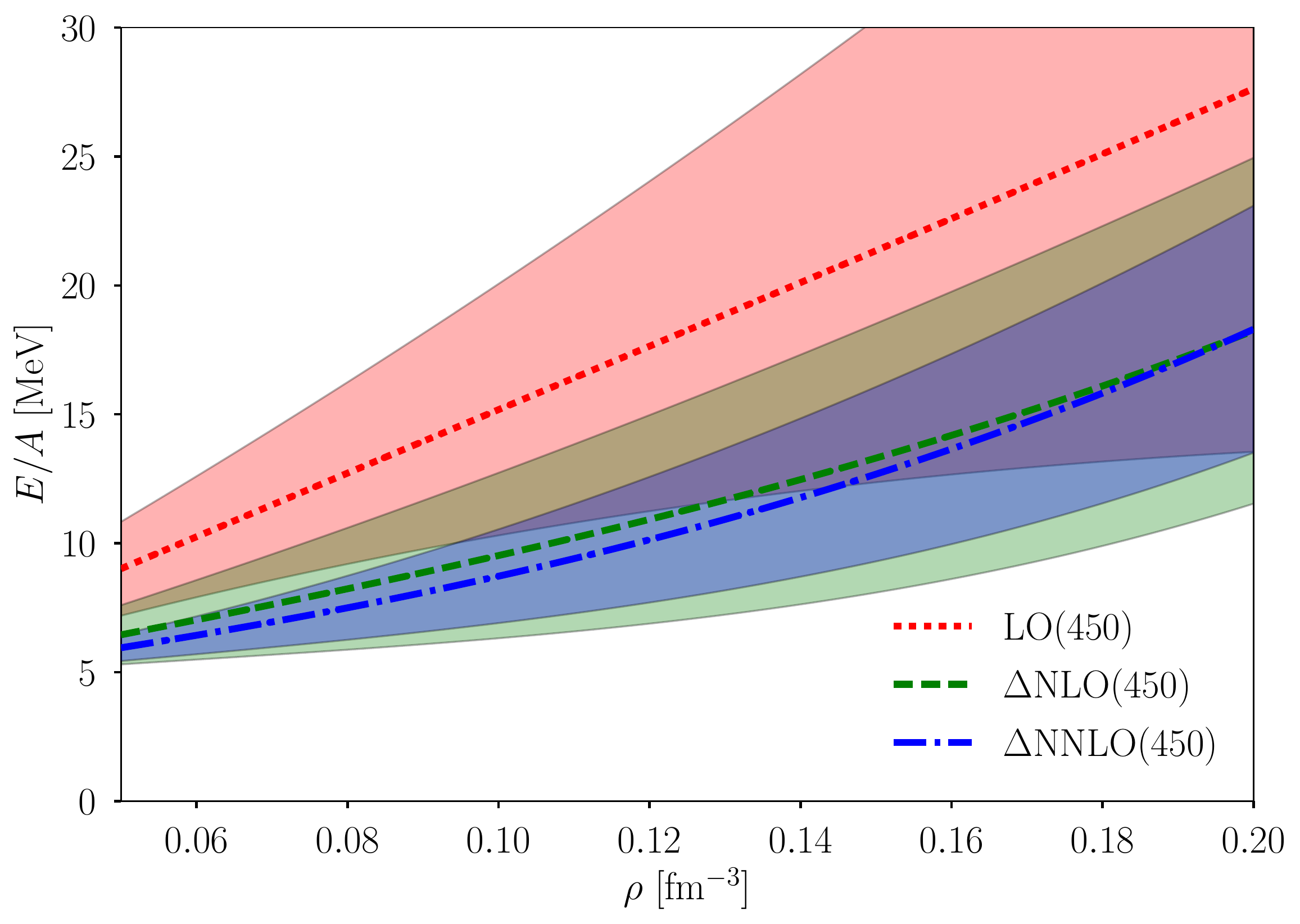}
\caption{(Color online) Energy per nucleon (in MeV) of symmetric
  nuclear matter (upper row) and pure neutron matter (lower row) up to
  third order in \xEFT{} without (left column) and with (right column)
  explicit inclusion of the $\Delta$-isobar in \xEFT{}. All
  interaction employ a momentum cutoff $\Lambda = 450$~MeV. Shaded
  areas indicate the estimated EFT-truncation errors, and the (square)
  diamond marks the saturation point in symmetric nuclear matter for
  ($\Delta$NLO) $\Delta$NNLO. The black rectangle indicates the
  region $E/A=-16\pm 0.5$ MeV and $\rho=0.16\pm0.01$ fm$^{-3}$.}
\label{nm450}
\end{figure*}
\begin{figure}[htb]
  \includegraphics[width=0.45\textwidth]{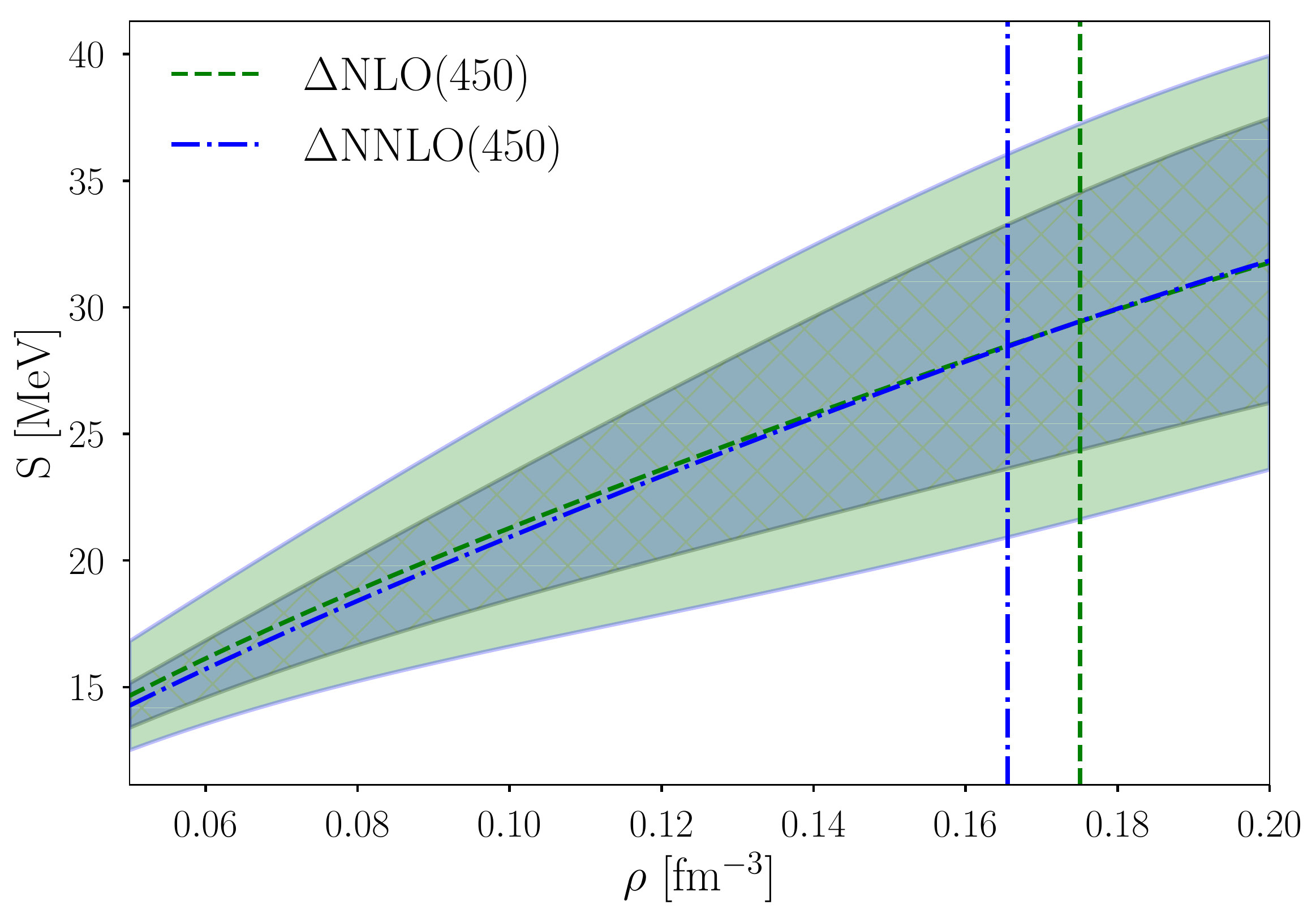}
  \caption{(Color online) The symmetry energy as a function of the
    density for $\Delta$NLO (green), and $\Delta$NNLO (blue) with
    $\Delta$-full interaction at cutoff of 450~MeV, with uncertainties
    shown as shaded areas.}
  \label{sym_energy}
  \end{figure}

In Fig.~\ref{nm450} we compare the results for the energy per nucleon
at different densities in symmetric nuclear matter and pure neutron
matter using $\Delta$-full and $\Delta$-less interactions with a
momentum cutoff $\Lambda=$450 MeV at LO, NLO, and NNLO.  The
saturation points in symmetric matter at NLO and NNLO shift towards
considerably more realistic values upon inclusion of the
$\Delta$. This observation is consistent with our results for finite
nuclei. For the EFT truncation uncertainty we use
Eq.~\ref{eq:eft_error}, a relevant momentum scale $p=p_F$, and the
breakdown momentum $\Lambda_b=500$~MeV. The uncertainties also make
the accelerated convergence and consistency of the $\Delta$-full
expansion more apparent. We note that our breakdown scale $\Lambda_b$
is rather conservative. For $\Lambda_b\gtrsim 650$~MeV the
truncation-error bands of the $\Delta$-full and $\Delta$-less
NNLO(450) interactions no longer overlap in the region of the
empirical saturation density.  We also note that nuclear matter does
not saturate at LO in the range of densities we studied, and we remind
the reader once more that the $\Delta$ does not enter at this chiral
order.


%
%

Figure~\ref{sym_energy} shows the difference between the equations of
state for neutron matter and symmetric nuclear matter for the
$\Lambda=450$~MeV cutoff. At the saturation density ($\rho_0$),
indicated as vertical lines for the different orders, this yields the
symmetry energy ($S_0$).  Our results are $\rho_0= 0.18(1)$~fm$^{-3}$,
$22.8\lesssim S_0\lesssim 36.5$~MeV, and $46\lesssim L \lesssim
65$~MeV at $\Delta$NLO, and $\rho_0=0.165(1)$~fm$^{-3}$, $23.6\lesssim
S_0\lesssim 33.3$~MeV, and $32\lesssim L \lesssim 67$~MeV at
$\Delta$NNLO. The estimated EFT truncation error for $\rho_0$ is very
small at $\Delta$NNLO because its central value and lower and upper
bounds have essentially the same saturation point.
The estimated EFT truncation error for $S_0$ is the maximum difference
between the energies per particle in neutron matter and symmetric
nuclear matter, at the saturation point.  This uncertainty also
decreases with increasing order. Finally, the estimated uncertainty in
the slope ($L$) of the symmetry energy is taken from the ranges of
slopes of $S_0$ at its upper and lower values.  It is large even at
$\Delta$NNLO and reflects that the slope in neutron matter exhibits a
greater variance at $\Delta$NNLO than at $\Delta$NLO, see
Fig.~\ref{nm450}. We note that our predictions for the symmetry energy
and its density derivative at $\Delta$NLO and $\Delta$NNLO are
consistent with the recent estimates of Ref.~\cite{tsang2012,tews2016}.

\section{Summary}
We presented results for selected finite nuclei and
infinite nucleonic matter using optimized interactions from \xEFT{}
with explicit $\Delta$-isobar degree-of-freedom. We optimized both
$\Delta$-full and $\Delta$-less interactions order-by-order in the
power counting up to NNLO, for two different cutoffs, and with $\pi N$
LECs from a recent Roy-Steiner analysis of $\pi N$ scattering. The
$NN$ contact potentials up to NNLO were adjusted to $NN$ phase shifts,
while the short-ranged parts of the $NNN$ interactions were
constrained by energy and radius data on $^{4}$He. We emphasize that
the only differences between the $\Delta$-full and $\Delta$-less
interactions are due to the explicit inclusion of the $\Delta$
isobar. In a detailed comparison, we found that radii in nuclei up to
$^{48}$Ca are accurate within EFT-truncation error estimates, and that
binding energies -- while improving order-by-order in precision --
somewhat underbind heavier nuclei. The saturation point in nuclear
matter is consistent with data within EFT error estimates. Our
results also show that the inclusion of $\Delta$-isobars in the
nuclear interaction can address the long-standing problem regarding
nuclear saturation. This work therefore provides a valuable starting
point for constructing more refined $\Delta$-full \xEFT{}
interactions, also at higher chiral orders, with improved uncertainty
estimates.

\begin{acknowledgments}
  We would like to thank Hermann Krebs for valuable discussions and
  input on the manuscript, Kai Hebeler for providing us with matrix
  elements in Jacobi coordinates for the three-nucleon interaction,
  and Gustav Jansen for providing us with the code that transforms
  three-nucleon matrix elements to the laboratory system. This
  material is based upon work supported by the Swedish Research
  Council under Grant No. 2015- 00225, and the Marie Sklodowska Curie
  Actions, Cofund, Project INCA 600398, the U.S. Department of Energy,
  Office of Science, Office of Nuclear Physics under Award Numbers
  DEFG02-96ER40963 (University of Tennessee), DE-SC0008499 and
  DE-SC0018223 (NUCLEI SciDAC collaboration), and the Field Work
  Proposal ERKBP57 at Oak Ridge National Laboratory. Computer time was
  provided by the Innovative and Novel Computational Impact on Theory
  and Experiment (INCITE) program. This research used resources of the
  Oak Ridge Leadership Computing Facility located in the Oak Ridge
  National Laboratory, which is supported by the Office of Science of
  the Department of Energy under Contract No.  DE-AC05-00OR22725, and
  used computational resources of the National Center for
  Computational Sciences, the National Institute for Computational
  Sciences, the Swedish National Infrastructure for Computing (SNIC).
\end{acknowledgments}

\bibliography{refs,refs3}

\begin{thebibliography}{76}%
\makeatletter
\providecommand \@ifxundefined [1]{%
 \@ifx{#1\undefined}
}%
\providecommand \@ifnum [1]{%
 \ifnum #1\expandafter \@firstoftwo
 \else \expandafter \@secondoftwo
 \fi
}%
\providecommand \@ifx [1]{%
 \ifx #1\expandafter \@firstoftwo
 \else \expandafter \@secondoftwo
 \fi
}%
\providecommand \natexlab [1]{#1}%
\providecommand \enquote  [1]{``#1''}%
\providecommand \bibnamefont  [1]{#1}%
\providecommand \bibfnamefont [1]{#1}%
\providecommand \citenamefont [1]{#1}%
\providecommand \href@noop [0]{\@secondoftwo}%
\providecommand \href [0]{\begingroup \@sanitize@url \@href}%
\providecommand \@href[1]{\@@startlink{#1}\@@href}%
\providecommand \@@href[1]{\endgroup#1\@@endlink}%
\providecommand \@sanitize@url [0]{\catcode `\\12\catcode `\$12\catcode
  `\&12\catcode `\#12\catcode `\^12\catcode `\_12\catcode `\%12\relax}%
\providecommand \@@startlink[1]{}%
\providecommand \@@endlink[0]{}%
\providecommand \url  [0]{\begingroup\@sanitize@url \@url }%
\providecommand \@url [1]{\endgroup\@href {#1}{\urlprefix }}%
\providecommand \urlprefix  [0]{URL }%
\providecommand \Eprint [0]{\href }%
\providecommand \doibase [0]{http://dx.doi.org/}%
\providecommand \selectlanguage [0]{\@gobble}%
\providecommand \bibinfo  [0]{\@secondoftwo}%
\providecommand \bibfield  [0]{\@secondoftwo}%
\providecommand \translation [1]{[#1]}%
\providecommand \BibitemOpen [0]{}%
\providecommand \bibitemStop [0]{}%
\providecommand \bibitemNoStop [0]{.\EOS\space}%
\providecommand \EOS [0]{\spacefactor3000\relax}%
\providecommand \BibitemShut  [1]{\csname bibitem#1\endcsname}%
\let\auto@bib@innerbib\@empty
\bibitem [{\citenamefont {Kamada}\ \emph {et~al.}(2001)\citenamefont {Kamada},
  \citenamefont {Nogga}, \citenamefont {Gl\"ockle}, \citenamefont {Hiyama},
  \citenamefont {Kamimura}, \citenamefont {Varga}, \citenamefont {Suzuki},
  \citenamefont {Viviani}, \citenamefont {Kievsky}, \citenamefont {Rosati},
  \citenamefont {Carlson}, \citenamefont {Pieper}, \citenamefont {Wiringa},
  \citenamefont {Navr\'atil}, \citenamefont {Barrett}, \citenamefont {Barnea},
  \citenamefont {Leidemann},\ and\ \citenamefont {Orlandini}}]{kamada2001}%
  \BibitemOpen
  \bibfield  {author} {\bibinfo {author} {\bibfnamefont {H.}~\bibnamefont
  {Kamada}}, \bibinfo {author} {\bibfnamefont {A.}~\bibnamefont {Nogga}},
  \bibinfo {author} {\bibfnamefont {W.}~\bibnamefont {Gl\"ockle}}, \bibinfo
  {author} {\bibfnamefont {E.}~\bibnamefont {Hiyama}}, \bibinfo {author}
  {\bibfnamefont {M.}~\bibnamefont {Kamimura}}, \bibinfo {author}
  {\bibfnamefont {K.}~\bibnamefont {Varga}}, \bibinfo {author} {\bibfnamefont
  {Y.}~\bibnamefont {Suzuki}}, \bibinfo {author} {\bibfnamefont
  {M.}~\bibnamefont {Viviani}}, \bibinfo {author} {\bibfnamefont
  {A.}~\bibnamefont {Kievsky}}, \bibinfo {author} {\bibfnamefont
  {S.}~\bibnamefont {Rosati}}, \bibinfo {author} {\bibfnamefont
  {J.}~\bibnamefont {Carlson}}, \bibinfo {author} {\bibfnamefont {Steven~C.}\
  \bibnamefont {Pieper}}, \bibinfo {author} {\bibfnamefont {R.~B.}\
  \bibnamefont {Wiringa}}, \bibinfo {author} {\bibfnamefont {P.}~\bibnamefont
  {Navr\'atil}}, \bibinfo {author} {\bibfnamefont {B.~R.}\ \bibnamefont
  {Barrett}}, \bibinfo {author} {\bibfnamefont {N.}~\bibnamefont {Barnea}},
  \bibinfo {author} {\bibfnamefont {W.}~\bibnamefont {Leidemann}}, \ and\
  \bibinfo {author} {\bibfnamefont {G.}~\bibnamefont {Orlandini}},\ }\bibfield
  {title} {\enquote {\bibinfo {title} {Benchmark test calculation of a
  four-nucleon bound state},}\ }\href {\doibase 10.1103/PhysRevC.64.044001}
  {\bibfield  {journal} {\bibinfo  {journal} {Phys. Rev. C}\ }\textbf {\bibinfo
  {volume} {64}},\ \bibinfo {pages} {044001} (\bibinfo {year}
  {2001})}\BibitemShut {NoStop}%
\bibitem [{\citenamefont {Pieper}\ and\ \citenamefont
  {Wiringa}(2001)}]{pieper2001}%
  \BibitemOpen
  \bibfield  {author} {\bibinfo {author} {\bibfnamefont {S.~C.}\ \bibnamefont
  {Pieper}}\ and\ \bibinfo {author} {\bibfnamefont {R.~B.}\ \bibnamefont
  {Wiringa}},\ }\bibfield  {title} {\enquote {\bibinfo {title} {{Quantum Monte
  Carlo calculations of light nuclei}},}\ }\href {\doibase
  10.1146/annurev.nucl.51.101701.132506} {\bibfield  {journal} {\bibinfo
  {journal} {Ann. Rev. Nucl. Part. Sci.}\ }\textbf {\bibinfo {volume} {51}},\
  \bibinfo {pages} {53--90} (\bibinfo {year} {2001})}\BibitemShut {NoStop}%
\bibitem [{\citenamefont {Navr{\'a}til}\ \emph {et~al.}(2009)\citenamefont
  {Navr{\'a}til}, \citenamefont {Quaglioni}, \citenamefont {Stetcu},\ and\
  \citenamefont {Barrett}}]{navratil2009}%
  \BibitemOpen
  \bibfield  {author} {\bibinfo {author} {\bibfnamefont {P.}~\bibnamefont
  {Navr{\'a}til}}, \bibinfo {author} {\bibfnamefont {S.}~\bibnamefont
  {Quaglioni}}, \bibinfo {author} {\bibfnamefont {I.}~\bibnamefont {Stetcu}}, \
  and\ \bibinfo {author} {\bibfnamefont {B.~R.}\ \bibnamefont {Barrett}},\
  }\bibfield  {title} {\enquote {\bibinfo {title} {Recent developments in
  no-core shell-model calculations},}\ }\href
  {http://stacks.iop.org/0954-3899/36/i=8/a=083101} {\bibfield  {journal}
  {\bibinfo  {journal} {J. Phys. G: Nucl. Part. Phys.}\ }\textbf {\bibinfo
  {volume} {36}},\ \bibinfo {pages} {083101} (\bibinfo {year}
  {2009})}\BibitemShut {NoStop}%
\bibitem [{\citenamefont {Barrett}\ \emph {et~al.}(2013)\citenamefont
  {Barrett}, \citenamefont {Navr{\'a}til},\ and\ \citenamefont
  {Vary}}]{barrett2013}%
  \BibitemOpen
  \bibfield  {author} {\bibinfo {author} {\bibfnamefont {Bruce~R.}\
  \bibnamefont {Barrett}}, \bibinfo {author} {\bibfnamefont {Petr}\
  \bibnamefont {Navr{\'a}til}}, \ and\ \bibinfo {author} {\bibfnamefont
  {James~P.}\ \bibnamefont {Vary}},\ }\bibfield  {title} {\enquote {\bibinfo
  {title} {Ab initio no core shell model},}\ }\href {\doibase
  10.1016/j.ppnp.2012.10.003} {\bibfield  {journal} {\bibinfo  {journal} {Prog.
  Part. Nucl. Phys.}\ }\textbf {\bibinfo {volume} {69}},\ \bibinfo {pages} {131
  -- 181} (\bibinfo {year} {2013})}\BibitemShut {NoStop}%
\bibitem [{\citenamefont {Hagen}\ \emph
  {et~al.}(2014{\natexlab{a}})\citenamefont {Hagen}, \citenamefont
  {Papenbrock}, \citenamefont {Hjorth-Jensen},\ and\ \citenamefont
  {Dean}}]{hagen2014}%
  \BibitemOpen
  \bibfield  {author} {\bibinfo {author} {\bibfnamefont {G.}~\bibnamefont
  {Hagen}}, \bibinfo {author} {\bibfnamefont {T.}~\bibnamefont {Papenbrock}},
  \bibinfo {author} {\bibfnamefont {M.}~\bibnamefont {Hjorth-Jensen}}, \ and\
  \bibinfo {author} {\bibfnamefont {D.~J.}\ \bibnamefont {Dean}},\ }\bibfield
  {title} {\enquote {\bibinfo {title} {Coupled-cluster computations of atomic
  nuclei},}\ }\href {\doibase 10.1088/0034-4885/77/9/096302} {\bibfield
  {journal} {\bibinfo  {journal} {Rep. Prog. Phys.}\ }\textbf {\bibinfo
  {volume} {77}},\ \bibinfo {pages} {096302} (\bibinfo {year}
  {2014}{\natexlab{a}})}\BibitemShut {NoStop}%
\bibitem [{\citenamefont {L{\"a}hde}\ \emph {et~al.}(2014)\citenamefont
  {L{\"a}hde}, \citenamefont {Epelbaum}, \citenamefont {Krebs}, \citenamefont
  {Lee}, \citenamefont {Mei\ss{}ner},\ and\ \citenamefont {Rupak}}]{lahde2014}%
  \BibitemOpen
  \bibfield  {author} {\bibinfo {author} {\bibfnamefont {T.~A.}\ \bibnamefont
  {L{\"a}hde}}, \bibinfo {author} {\bibfnamefont {E.}~\bibnamefont {Epelbaum}},
  \bibinfo {author} {\bibfnamefont {H.}~\bibnamefont {Krebs}}, \bibinfo
  {author} {\bibfnamefont {D.}~\bibnamefont {Lee}}, \bibinfo {author}
  {\bibfnamefont {U.-G.}\ \bibnamefont {Mei\ss{}ner}}, \ and\ \bibinfo {author}
  {\bibfnamefont {G.}~\bibnamefont {Rupak}},\ }\bibfield  {title} {\enquote
  {\bibinfo {title} {Lattice effective field theory for medium-mass nuclei},}\
  }\href {\doibase 10.1016/j.physletb.2014.03.023} {\bibfield  {journal}
  {\bibinfo  {journal} {Phys. Lett. B}\ }\textbf {\bibinfo {volume} {732}},\
  \bibinfo {pages} {110 -- 115} (\bibinfo {year} {2014})}\BibitemShut {NoStop}%
\bibitem [{\citenamefont {{Hagen}}\ \emph {et~al.}(2016)\citenamefont
  {{Hagen}}, \citenamefont {{Ekstr{\"o}m}}, \citenamefont {{Forss{\'e}n}},
  \citenamefont {{Jansen}}, \citenamefont {{Nazarewicz}}, \citenamefont
  {{Papenbrock}}, \citenamefont {{Wendt}}, \citenamefont {{Bacca}},
  \citenamefont {{Barnea}}, \citenamefont {{Carlsson}}, \citenamefont
  {{Drischler}}, \citenamefont {{Hebeler}}, \citenamefont {{Hjorth-Jensen}},
  \citenamefont {{Miorelli}}, \citenamefont {{Orlandini}}, \citenamefont
  {{Schwenk}},\ and\ \citenamefont {{Simonis}}}]{hagen2015}%
  \BibitemOpen
  \bibfield  {author} {\bibinfo {author} {\bibfnamefont {G.}~\bibnamefont
  {{Hagen}}}, \bibinfo {author} {\bibfnamefont {A.}~\bibnamefont
  {{Ekstr{\"o}m}}}, \bibinfo {author} {\bibfnamefont {C.}~\bibnamefont
  {{Forss{\'e}n}}}, \bibinfo {author} {\bibfnamefont {G.~R.}\ \bibnamefont
  {{Jansen}}}, \bibinfo {author} {\bibfnamefont {W.}~\bibnamefont
  {{Nazarewicz}}}, \bibinfo {author} {\bibfnamefont {T.}~\bibnamefont
  {{Papenbrock}}}, \bibinfo {author} {\bibfnamefont {K.~A.}\ \bibnamefont
  {{Wendt}}}, \bibinfo {author} {\bibfnamefont {S.}~\bibnamefont {{Bacca}}},
  \bibinfo {author} {\bibfnamefont {N.}~\bibnamefont {{Barnea}}}, \bibinfo
  {author} {\bibfnamefont {B.}~\bibnamefont {{Carlsson}}}, \bibinfo {author}
  {\bibfnamefont {C.}~\bibnamefont {{Drischler}}}, \bibinfo {author}
  {\bibfnamefont {K.}~\bibnamefont {{Hebeler}}}, \bibinfo {author}
  {\bibfnamefont {M.}~\bibnamefont {{Hjorth-Jensen}}}, \bibinfo {author}
  {\bibfnamefont {M.}~\bibnamefont {{Miorelli}}}, \bibinfo {author}
  {\bibfnamefont {G.}~\bibnamefont {{Orlandini}}}, \bibinfo {author}
  {\bibfnamefont {A.}~\bibnamefont {{Schwenk}}}, \ and\ \bibinfo {author}
  {\bibfnamefont {J.}~\bibnamefont {{Simonis}}},\ }\bibfield  {title} {\enquote
  {\bibinfo {title} {{Neutron and weak-charge distributions of the $^{48}$Ca
  nucleus}},}\ }\href {\doibase 10.1038/nphys3529} {\bibfield  {journal}
  {\bibinfo  {journal} {Nature Physics}\ }\textbf {\bibinfo {volume} {12}},\
  \bibinfo {pages} {186} (\bibinfo {year} {2016})}\BibitemShut {NoStop}%
\bibitem [{\citenamefont {Hergert}\ \emph {et~al.}(2016)\citenamefont
  {Hergert}, \citenamefont {Bogner}, \citenamefont {Morris}, \citenamefont
  {Schwenk},\ and\ \citenamefont {Tsukiyama}}]{hergert2016}%
  \BibitemOpen
  \bibfield  {author} {\bibinfo {author} {\bibfnamefont {H.}~\bibnamefont
  {Hergert}}, \bibinfo {author} {\bibfnamefont {S.~K.}\ \bibnamefont {Bogner}},
  \bibinfo {author} {\bibfnamefont {T.~D.}\ \bibnamefont {Morris}}, \bibinfo
  {author} {\bibfnamefont {A.}~\bibnamefont {Schwenk}}, \ and\ \bibinfo
  {author} {\bibfnamefont {K.}~\bibnamefont {Tsukiyama}},\ }\bibfield  {title}
  {\enquote {\bibinfo {title} {The in-medium similarity renormalization group:
  A novel ab initio method for nuclei},}\ }\href {\doibase
  10.1016/j.physrep.2015.12.007} {\bibfield  {journal} {\bibinfo  {journal}
  {Phys. Rep.}\ }\textbf {\bibinfo {volume} {621}},\ \bibinfo {pages} {165 --
  222} (\bibinfo {year} {2016})}\BibitemShut {NoStop}%
\bibitem [{\citenamefont {Hagen}\ \emph {et~al.}(2016)\citenamefont {Hagen},
  \citenamefont {Jansen},\ and\ \citenamefont {Papenbrock}}]{hagen2016b}%
  \BibitemOpen
  \bibfield  {author} {\bibinfo {author} {\bibfnamefont {G.}~\bibnamefont
  {Hagen}}, \bibinfo {author} {\bibfnamefont {G.~R.}\ \bibnamefont {Jansen}}, \
  and\ \bibinfo {author} {\bibfnamefont {T.}~\bibnamefont {Papenbrock}},\
  }\bibfield  {title} {\enquote {\bibinfo {title} {Structure of
  $^{78}\mathrm{Ni}$ from first-principles computations},}\ }\href {\doibase
  10.1103/PhysRevLett.117.172501} {\bibfield  {journal} {\bibinfo  {journal}
  {Phys. Rev. Lett.}\ }\textbf {\bibinfo {volume} {117}},\ \bibinfo {pages}
  {172501} (\bibinfo {year} {2016})}\BibitemShut {NoStop}%
\bibitem [{\citenamefont {Epelbaum}\ \emph {et~al.}(2000)\citenamefont
  {Epelbaum}, \citenamefont {Gl{\"o}ckle},\ and\ \citenamefont
  {Mei{\ss}ner}}]{epelbaum2000}%
  \BibitemOpen
  \bibfield  {author} {\bibinfo {author} {\bibfnamefont {E.}~\bibnamefont
  {Epelbaum}}, \bibinfo {author} {\bibfnamefont {W.}~\bibnamefont
  {Gl{\"o}ckle}}, \ and\ \bibinfo {author} {\bibfnamefont {U.-G.}\ \bibnamefont
  {Mei{\ss}ner}},\ }\bibfield  {title} {\enquote {\bibinfo {title} {Nuclear
  forces from chiral lagrangians using the method of unitary transformation ii:
  The two-nucleon system},}\ }\href {\doibase 10.1016/S0375-9474(99)00821-0}
  {\bibfield  {journal} {\bibinfo  {journal} {Nucl. Phys. A}\ }\textbf
  {\bibinfo {volume} {671}},\ \bibinfo {pages} {295 -- 331} (\bibinfo {year}
  {2000})}\BibitemShut {NoStop}%
\bibitem [{\citenamefont {Epelbaum}\ \emph {et~al.}(2002)\citenamefont
  {Epelbaum}, \citenamefont {Nogga}, \citenamefont {Gl\"ockle}, \citenamefont
  {Kamada}, \citenamefont {Mei\ss{}ner},\ and\ \citenamefont
  {Wita\l{}a}}]{epelbaum2002}%
  \BibitemOpen
  \bibfield  {author} {\bibinfo {author} {\bibfnamefont {E.}~\bibnamefont
  {Epelbaum}}, \bibinfo {author} {\bibfnamefont {A.}~\bibnamefont {Nogga}},
  \bibinfo {author} {\bibfnamefont {W.}~\bibnamefont {Gl\"ockle}}, \bibinfo
  {author} {\bibfnamefont {H.}~\bibnamefont {Kamada}}, \bibinfo {author}
  {\bibfnamefont {U.-G.}\ \bibnamefont {Mei\ss{}ner}}, \ and\ \bibinfo {author}
  {\bibfnamefont {H.}~\bibnamefont {Wita\l{}a}},\ }\bibfield  {title} {\enquote
  {\bibinfo {title} {Three-nucleon forces from chiral effective field
  theory},}\ }\href {\doibase 10.1103/PhysRevC.66.064001} {\bibfield  {journal}
  {\bibinfo  {journal} {Phys. Rev. C}\ }\textbf {\bibinfo {volume} {66}},\
  \bibinfo {pages} {064001} (\bibinfo {year} {2002})}\BibitemShut {NoStop}%
\bibitem [{\citenamefont {Entem}\ and\ \citenamefont
  {Machleidt}(2003)}]{entem2003}%
  \BibitemOpen
  \bibfield  {author} {\bibinfo {author} {\bibfnamefont {D.~R.}\ \bibnamefont
  {Entem}}\ and\ \bibinfo {author} {\bibfnamefont {R.}~\bibnamefont
  {Machleidt}},\ }\bibfield  {title} {\enquote {\bibinfo {title} {Accurate
  charge-dependent nucleon-nucleon potential at fourth order of chiral
  perturbation theory},}\ }\href {\doibase 10.1103/PhysRevC.68.041001}
  {\bibfield  {journal} {\bibinfo  {journal} {Phys. Rev. C}\ }\textbf {\bibinfo
  {volume} {68}},\ \bibinfo {pages} {041001} (\bibinfo {year}
  {2003})}\BibitemShut {NoStop}%
\bibitem [{\citenamefont {Shirokov}\ \emph {et~al.}(2004)\citenamefont
  {Shirokov}, \citenamefont {Mazur}, \citenamefont {Zaytsev}, \citenamefont
  {Vary},\ and\ \citenamefont {Weber}}]{shirokov2004}%
  \BibitemOpen
  \bibfield  {author} {\bibinfo {author} {\bibfnamefont {A.~M.}\ \bibnamefont
  {Shirokov}}, \bibinfo {author} {\bibfnamefont {A.~I.}\ \bibnamefont {Mazur}},
  \bibinfo {author} {\bibfnamefont {S.~A.}\ \bibnamefont {Zaytsev}}, \bibinfo
  {author} {\bibfnamefont {J.~P.}\ \bibnamefont {Vary}}, \ and\ \bibinfo
  {author} {\bibfnamefont {T.~A.}\ \bibnamefont {Weber}},\ }\bibfield  {title}
  {\enquote {\bibinfo {title} {Nucleon-nucleon interaction in the $j$-matrix
  inverse scattering approach and few-nucleon systems},}\ }\href {\doibase
  10.1103/PhysRevC.70.044005} {\bibfield  {journal} {\bibinfo  {journal} {Phys.
  Rev. C}\ }\textbf {\bibinfo {volume} {70}},\ \bibinfo {pages} {044005}
  (\bibinfo {year} {2004})}\BibitemShut {NoStop}%
\bibitem [{\citenamefont {Hebeler}\ \emph {et~al.}(2011)\citenamefont
  {Hebeler}, \citenamefont {Bogner}, \citenamefont {Furnstahl}, \citenamefont
  {Nogga},\ and\ \citenamefont {Schwenk}}]{hebeler2011}%
  \BibitemOpen
  \bibfield  {author} {\bibinfo {author} {\bibfnamefont {K.}~\bibnamefont
  {Hebeler}}, \bibinfo {author} {\bibfnamefont {S.~K.}\ \bibnamefont {Bogner}},
  \bibinfo {author} {\bibfnamefont {R.~J.}\ \bibnamefont {Furnstahl}}, \bibinfo
  {author} {\bibfnamefont {A.}~\bibnamefont {Nogga}}, \ and\ \bibinfo {author}
  {\bibfnamefont {A.}~\bibnamefont {Schwenk}},\ }\bibfield  {title} {\enquote
  {\bibinfo {title} {Improved nuclear matter calculations from chiral
  low-momentum interactions},}\ }\href {\doibase 10.1103/PhysRevC.83.031301}
  {\bibfield  {journal} {\bibinfo  {journal} {Phys. Rev. C}\ }\textbf {\bibinfo
  {volume} {83}},\ \bibinfo {pages} {031301} (\bibinfo {year}
  {2011})}\BibitemShut {NoStop}%
\bibitem [{\citenamefont {Ekstr\"om}\ \emph {et~al.}(2013)\citenamefont
  {Ekstr\"om}, \citenamefont {Baardsen}, \citenamefont {Forss\'en},
  \citenamefont {Hagen}, \citenamefont {Hjorth-Jensen}, \citenamefont {Jansen},
  \citenamefont {Machleidt}, \citenamefont {Nazarewicz}, \citenamefont
  {Papenbrock}, \citenamefont {Sarich},\ and\ \citenamefont
  {Wild}}]{ekstrom2013}%
  \BibitemOpen
  \bibfield  {author} {\bibinfo {author} {\bibfnamefont {A.}~\bibnamefont
  {Ekstr\"om}}, \bibinfo {author} {\bibfnamefont {G.}~\bibnamefont {Baardsen}},
  \bibinfo {author} {\bibfnamefont {C.}~\bibnamefont {Forss\'en}}, \bibinfo
  {author} {\bibfnamefont {G.}~\bibnamefont {Hagen}}, \bibinfo {author}
  {\bibfnamefont {M.}~\bibnamefont {Hjorth-Jensen}}, \bibinfo {author}
  {\bibfnamefont {G.~R.}\ \bibnamefont {Jansen}}, \bibinfo {author}
  {\bibfnamefont {R.}~\bibnamefont {Machleidt}}, \bibinfo {author}
  {\bibfnamefont {W.}~\bibnamefont {Nazarewicz}}, \bibinfo {author}
  {\bibfnamefont {T.}~\bibnamefont {Papenbrock}}, \bibinfo {author}
  {\bibfnamefont {J.}~\bibnamefont {Sarich}}, \ and\ \bibinfo {author}
  {\bibfnamefont {S.~M.}\ \bibnamefont {Wild}},\ }\bibfield  {title} {\enquote
  {\bibinfo {title} {Optimized chiral nucleon-nucleon interaction at
  next-to-next-to-leading order},}\ }\href {\doibase
  10.1103/PhysRevLett.110.192502} {\bibfield  {journal} {\bibinfo  {journal}
  {Phys. Rev. Lett.}\ }\textbf {\bibinfo {volume} {110}},\ \bibinfo {pages}
  {192502} (\bibinfo {year} {2013})}\BibitemShut {NoStop}%
\bibitem [{\citenamefont {Entem}\ \emph {et~al.}(2015)\citenamefont {Entem},
  \citenamefont {Kaiser}, \citenamefont {Machleidt},\ and\ \citenamefont
  {Nosyk}}]{entem2015}%
  \BibitemOpen
  \bibfield  {author} {\bibinfo {author} {\bibfnamefont {D.~R.}\ \bibnamefont
  {Entem}}, \bibinfo {author} {\bibfnamefont {N.}~\bibnamefont {Kaiser}},
  \bibinfo {author} {\bibfnamefont {R.}~\bibnamefont {Machleidt}}, \ and\
  \bibinfo {author} {\bibfnamefont {Y.}~\bibnamefont {Nosyk}},\ }\bibfield
  {title} {\enquote {\bibinfo {title} {Peripheral nucleon-nucleon scattering at
  fifth order of chiral perturbation theory},}\ }\href {\doibase
  10.1103/PhysRevC.91.014002} {\bibfield  {journal} {\bibinfo  {journal} {Phys.
  Rev. C}\ }\textbf {\bibinfo {volume} {91}},\ \bibinfo {pages} {014002}
  (\bibinfo {year} {2015})}\BibitemShut {NoStop}%
\bibitem [{\citenamefont {Epelbaum}\ \emph
  {et~al.}(2015{\natexlab{a}})\citenamefont {Epelbaum}, \citenamefont {Krebs},\
  and\ \citenamefont {Mei\ss{}ner}}]{epelbaum2015}%
  \BibitemOpen
  \bibfield  {author} {\bibinfo {author} {\bibfnamefont {E.}~\bibnamefont
  {Epelbaum}}, \bibinfo {author} {\bibfnamefont {H.}~\bibnamefont {Krebs}}, \
  and\ \bibinfo {author} {\bibfnamefont {U.-G.}\ \bibnamefont {Mei\ss{}ner}},\
  }\bibfield  {title} {\enquote {\bibinfo {title} {Precision nucleon-nucleon
  potential at fifth order in the chiral expansion},}\ }\href {\doibase
  10.1103/PhysRevLett.115.122301} {\bibfield  {journal} {\bibinfo  {journal}
  {Phys. Rev. Lett.}\ }\textbf {\bibinfo {volume} {115}},\ \bibinfo {pages}
  {122301} (\bibinfo {year} {2015}{\natexlab{a}})}\BibitemShut {NoStop}%
\bibitem [{\citenamefont {Ekstr\"om}\ \emph {et~al.}(2015)\citenamefont
  {Ekstr\"om}, \citenamefont {Jansen}, \citenamefont {Wendt}, \citenamefont
  {Hagen}, \citenamefont {Papenbrock}, \citenamefont {Carlsson}, \citenamefont
  {Forss\'en}, \citenamefont {Hjorth-Jensen}, \citenamefont {Navr\'atil},\ and\
  \citenamefont {Nazarewicz}}]{ekstrom2015a}%
  \BibitemOpen
  \bibfield  {author} {\bibinfo {author} {\bibfnamefont {A.}~\bibnamefont
  {Ekstr\"om}}, \bibinfo {author} {\bibfnamefont {G.~R.}\ \bibnamefont
  {Jansen}}, \bibinfo {author} {\bibfnamefont {K.~A.}\ \bibnamefont {Wendt}},
  \bibinfo {author} {\bibfnamefont {G.}~\bibnamefont {Hagen}}, \bibinfo
  {author} {\bibfnamefont {T.}~\bibnamefont {Papenbrock}}, \bibinfo {author}
  {\bibfnamefont {B.~D.}\ \bibnamefont {Carlsson}}, \bibinfo {author}
  {\bibfnamefont {C.}~\bibnamefont {Forss\'en}}, \bibinfo {author}
  {\bibfnamefont {M.}~\bibnamefont {Hjorth-Jensen}}, \bibinfo {author}
  {\bibfnamefont {P.}~\bibnamefont {Navr\'atil}}, \ and\ \bibinfo {author}
  {\bibfnamefont {W.}~\bibnamefont {Nazarewicz}},\ }\bibfield  {title}
  {\enquote {\bibinfo {title} {Accurate nuclear radii and binding energies from
  a chiral interaction},}\ }\href {\doibase 10.1103/PhysRevC.91.051301}
  {\bibfield  {journal} {\bibinfo  {journal} {Phys. Rev. C}\ }\textbf {\bibinfo
  {volume} {91}},\ \bibinfo {pages} {051301} (\bibinfo {year}
  {2015})}\BibitemShut {NoStop}%
\bibitem [{\citenamefont {Lynn}\ \emph {et~al.}(2016)\citenamefont {Lynn},
  \citenamefont {Tews}, \citenamefont {Carlson}, \citenamefont {Gandolfi},
  \citenamefont {Gezerlis}, \citenamefont {Schmidt},\ and\ \citenamefont
  {Schwenk}}]{lynn2016}%
  \BibitemOpen
  \bibfield  {author} {\bibinfo {author} {\bibfnamefont {J.~E.}\ \bibnamefont
  {Lynn}}, \bibinfo {author} {\bibfnamefont {I.}~\bibnamefont {Tews}}, \bibinfo
  {author} {\bibfnamefont {J.}~\bibnamefont {Carlson}}, \bibinfo {author}
  {\bibfnamefont {S.}~\bibnamefont {Gandolfi}}, \bibinfo {author}
  {\bibfnamefont {A.}~\bibnamefont {Gezerlis}}, \bibinfo {author}
  {\bibfnamefont {K.~E.}\ \bibnamefont {Schmidt}}, \ and\ \bibinfo {author}
  {\bibfnamefont {A.}~\bibnamefont {Schwenk}},\ }\bibfield  {title} {\enquote
  {\bibinfo {title} {Chiral three-nucleon interactions in light nuclei,
  neutron-$\ensuremath{\alpha}$ scattering, and neutron matter},}\ }\href
  {\doibase 10.1103/PhysRevLett.116.062501} {\bibfield  {journal} {\bibinfo
  {journal} {Phys. Rev. Lett.}\ }\textbf {\bibinfo {volume} {116}},\ \bibinfo
  {pages} {062501} (\bibinfo {year} {2016})}\BibitemShut {NoStop}%
\bibitem [{\citenamefont {Carlsson}\ \emph {et~al.}(2016)\citenamefont
  {Carlsson}, \citenamefont {Ekstr\"om}, \citenamefont {Forss\'en},
  \citenamefont {Str\"omberg}, \citenamefont {Jansen}, \citenamefont {Lilja},
  \citenamefont {Lindby}, \citenamefont {Mattsson},\ and\ \citenamefont
  {Wendt}}]{carlsson2016}%
  \BibitemOpen
  \bibfield  {author} {\bibinfo {author} {\bibfnamefont {B.~D.}\ \bibnamefont
  {Carlsson}}, \bibinfo {author} {\bibfnamefont {A.}~\bibnamefont {Ekstr\"om}},
  \bibinfo {author} {\bibfnamefont {C.}~\bibnamefont {Forss\'en}}, \bibinfo
  {author} {\bibfnamefont {D.~Fahlin}\ \bibnamefont {Str\"omberg}}, \bibinfo
  {author} {\bibfnamefont {G.~R.}\ \bibnamefont {Jansen}}, \bibinfo {author}
  {\bibfnamefont {O.}~\bibnamefont {Lilja}}, \bibinfo {author} {\bibfnamefont
  {M.}~\bibnamefont {Lindby}}, \bibinfo {author} {\bibfnamefont {B.~A.}\
  \bibnamefont {Mattsson}}, \ and\ \bibinfo {author} {\bibfnamefont {K.~A.}\
  \bibnamefont {Wendt}},\ }\bibfield  {title} {\enquote {\bibinfo {title}
  {Uncertainty analysis and order-by-order optimization of chiral nuclear
  interactions},}\ }\href {\doibase 10.1103/PhysRevX.6.011019} {\bibfield
  {journal} {\bibinfo  {journal} {Phys. Rev. X}\ }\textbf {\bibinfo {volume}
  {6}},\ \bibinfo {pages} {011019} (\bibinfo {year} {2016})}\BibitemShut
  {NoStop}%
\bibitem [{\citenamefont {Epelbaum}\ \emph {et~al.}(2009)\citenamefont
  {Epelbaum}, \citenamefont {Hammer},\ and\ \citenamefont
  {Mei\ss{}ner}}]{epelbaum2009}%
  \BibitemOpen
  \bibfield  {author} {\bibinfo {author} {\bibfnamefont {E.}~\bibnamefont
  {Epelbaum}}, \bibinfo {author} {\bibfnamefont {H.-W.}\ \bibnamefont
  {Hammer}}, \ and\ \bibinfo {author} {\bibfnamefont {U.-G.}\ \bibnamefont
  {Mei\ss{}ner}},\ }\bibfield  {title} {\enquote {\bibinfo {title} {Modern
  theory of nuclear forces},}\ }\href {\doibase 10.1103/RevModPhys.81.1773}
  {\bibfield  {journal} {\bibinfo  {journal} {Rev. Mod. Phys.}\ }\textbf
  {\bibinfo {volume} {81}},\ \bibinfo {pages} {1773--1825} (\bibinfo {year}
  {2009})}\BibitemShut {NoStop}%
\bibitem [{\citenamefont {Machleidt}\ and\ \citenamefont
  {Entem}(2011)}]{machleidt2011}%
  \BibitemOpen
  \bibfield  {author} {\bibinfo {author} {\bibfnamefont {R.}~\bibnamefont
  {Machleidt}}\ and\ \bibinfo {author} {\bibfnamefont {D.~R.}\ \bibnamefont
  {Entem}},\ }\bibfield  {title} {\enquote {\bibinfo {title} {Chiral effective
  field theory and nuclear forces},}\ }\href {\doibase
  10.1016/j.physrep.2011.02.001} {\bibfield  {journal} {\bibinfo  {journal}
  {Phys. Rep.}\ }\textbf {\bibinfo {volume} {503}},\ \bibinfo {pages} {1 -- 75}
  (\bibinfo {year} {2011})}\BibitemShut {NoStop}%
\bibitem [{\citenamefont {Binder}\ \emph {et~al.}(2014)\citenamefont {Binder},
  \citenamefont {Langhammer}, \citenamefont {Calci},\ and\ \citenamefont
  {Roth}}]{binder2013b}%
  \BibitemOpen
  \bibfield  {author} {\bibinfo {author} {\bibfnamefont {S.}~\bibnamefont
  {Binder}}, \bibinfo {author} {\bibfnamefont {J.}~\bibnamefont {Langhammer}},
  \bibinfo {author} {\bibfnamefont {A.}~\bibnamefont {Calci}}, \ and\ \bibinfo
  {author} {\bibfnamefont {R.}~\bibnamefont {Roth}},\ }\bibfield  {title}
  {\enquote {\bibinfo {title} {Ab initio path to heavy nuclei},}\ }\href
  {\doibase 10.1016/j.physletb.2014.07.010} {\bibfield  {journal} {\bibinfo
  {journal} {Phys. Lett. B}\ }\textbf {\bibinfo {volume} {736}},\ \bibinfo
  {pages} {119 -- 123} (\bibinfo {year} {2014})}\BibitemShut {NoStop}%
\bibitem [{\citenamefont {Simonis}\ \emph {et~al.}(2017)\citenamefont
  {Simonis}, \citenamefont {Stroberg}, \citenamefont {Hebeler}, \citenamefont
  {Holt},\ and\ \citenamefont {Schwenk}}]{simonis2017}%
  \BibitemOpen
  \bibfield  {author} {\bibinfo {author} {\bibfnamefont {J.}~\bibnamefont
  {Simonis}}, \bibinfo {author} {\bibfnamefont {S.~R.}\ \bibnamefont
  {Stroberg}}, \bibinfo {author} {\bibfnamefont {K.}~\bibnamefont {Hebeler}},
  \bibinfo {author} {\bibfnamefont {J.~D.}\ \bibnamefont {Holt}}, \ and\
  \bibinfo {author} {\bibfnamefont {A.}~\bibnamefont {Schwenk}},\ }\bibfield
  {title} {\enquote {\bibinfo {title} {Saturation with chiral interactions and
  consequences for finite nuclei},}\ }\href {\doibase
  10.1103/PhysRevC.96.014303} {\bibfield  {journal} {\bibinfo  {journal} {Phys.
  Rev. C}\ }\textbf {\bibinfo {volume} {96}},\ \bibinfo {pages} {014303}
  (\bibinfo {year} {2017})}\BibitemShut {NoStop}%
\bibitem [{\citenamefont {Elhatisari}\ \emph {et~al.}(2016)\citenamefont
  {Elhatisari}, \citenamefont {Li}, \citenamefont {Rokash}, \citenamefont
  {Alarc\'on}, \citenamefont {Du}, \citenamefont {Klein}, \citenamefont {Lu},
  \citenamefont {Mei\ss{}ner}, \citenamefont {Epelbaum}, \citenamefont {Krebs},
  \citenamefont {L\"ahde}, \citenamefont {Lee},\ and\ \citenamefont
  {Rupak}}]{elhatisari2016}%
  \BibitemOpen
  \bibfield  {author} {\bibinfo {author} {\bibfnamefont {Serdar}\ \bibnamefont
  {Elhatisari}}, \bibinfo {author} {\bibfnamefont {Ning}\ \bibnamefont {Li}},
  \bibinfo {author} {\bibfnamefont {Alexander}\ \bibnamefont {Rokash}},
  \bibinfo {author} {\bibfnamefont {Jose~Manuel}\ \bibnamefont {Alarc\'on}},
  \bibinfo {author} {\bibfnamefont {Dechuan}\ \bibnamefont {Du}}, \bibinfo
  {author} {\bibfnamefont {Nico}\ \bibnamefont {Klein}}, \bibinfo {author}
  {\bibfnamefont {Bing-nan}\ \bibnamefont {Lu}}, \bibinfo {author}
  {\bibfnamefont {Ulf-G.}\ \bibnamefont {Mei\ss{}ner}}, \bibinfo {author}
  {\bibfnamefont {Evgeny}\ \bibnamefont {Epelbaum}}, \bibinfo {author}
  {\bibfnamefont {Hermann}\ \bibnamefont {Krebs}}, \bibinfo {author}
  {\bibfnamefont {Timo~A.}\ \bibnamefont {L\"ahde}}, \bibinfo {author}
  {\bibfnamefont {Dean}\ \bibnamefont {Lee}}, \ and\ \bibinfo {author}
  {\bibfnamefont {Gautam}\ \bibnamefont {Rupak}},\ }\bibfield  {title}
  {\enquote {\bibinfo {title} {Nuclear binding near a quantum phase
  transition},}\ }\href {\doibase 10.1103/PhysRevLett.117.132501} {\bibfield
  {journal} {\bibinfo  {journal} {Phys. Rev. Lett.}\ }\textbf {\bibinfo
  {volume} {117}},\ \bibinfo {pages} {132501} (\bibinfo {year}
  {2016})}\BibitemShut {NoStop}%
\bibitem [{\citenamefont {van Kolck}(1994)}]{vankolck1994}%
  \BibitemOpen
  \bibfield  {author} {\bibinfo {author} {\bibfnamefont {U.}~\bibnamefont {van
  Kolck}},\ }\bibfield  {title} {\enquote {\bibinfo {title} {{Few-nucleon
  forces from chiral Lagrangians}},}\ }\href {\doibase
  10.1103/PhysRevC.49.2932} {\bibfield  {journal} {\bibinfo  {journal} {Phys.
  Rev. C}\ }\textbf {\bibinfo {volume} {49}},\ \bibinfo {pages} {2932--2941}
  (\bibinfo {year} {1994})}\BibitemShut {NoStop}%
\bibitem [{\citenamefont {Ord\'o\~nez}\ \emph {et~al.}(1994)\citenamefont
  {Ord\'o\~nez}, \citenamefont {Ray},\ and\ \citenamefont {van
  Kolck}}]{ordonez1994}%
  \BibitemOpen
  \bibfield  {author} {\bibinfo {author} {\bibfnamefont {C.}~\bibnamefont
  {Ord\'o\~nez}}, \bibinfo {author} {\bibfnamefont {L.}~\bibnamefont {Ray}}, \
  and\ \bibinfo {author} {\bibfnamefont {U.}~\bibnamefont {van Kolck}},\
  }\bibfield  {title} {\enquote {\bibinfo {title} {Nucleon-nucleon potential
  from an effective chiral lagrangian},}\ }\href {\doibase
  10.1103/PhysRevLett.72.1982} {\bibfield  {journal} {\bibinfo  {journal}
  {Phys. Rev. Lett.}\ }\textbf {\bibinfo {volume} {72}},\ \bibinfo {pages}
  {1982--1985} (\bibinfo {year} {1994})}\BibitemShut {NoStop}%
\bibitem [{\citenamefont {Ord\'o\~nez}\ \emph {et~al.}(1996)\citenamefont
  {Ord\'o\~nez}, \citenamefont {Ray},\ and\ \citenamefont {van
  Kolck}}]{ordonez1996}%
  \BibitemOpen
  \bibfield  {author} {\bibinfo {author} {\bibfnamefont {C.}~\bibnamefont
  {Ord\'o\~nez}}, \bibinfo {author} {\bibfnamefont {L.}~\bibnamefont {Ray}}, \
  and\ \bibinfo {author} {\bibfnamefont {U.}~\bibnamefont {van Kolck}},\
  }\bibfield  {title} {\enquote {\bibinfo {title} {Two-nucleon potential from
  chiral lagrangians},}\ }\href {\doibase 10.1103/PhysRevC.53.2086} {\bibfield
  {journal} {\bibinfo  {journal} {Phys. Rev. C}\ }\textbf {\bibinfo {volume}
  {53}},\ \bibinfo {pages} {2086--2105} (\bibinfo {year} {1996})}\BibitemShut
  {NoStop}%
\bibitem [{\citenamefont {Bernard}\ \emph {et~al.}(1997)\citenamefont
  {Bernard}, \citenamefont {Kaiser},\ and\ \citenamefont
  {Mei{\ss}ner}}]{Bernard1997}%
  \BibitemOpen
  \bibfield  {author} {\bibinfo {author} {\bibfnamefont {V.}~\bibnamefont
  {Bernard}}, \bibinfo {author} {\bibfnamefont {N.}~\bibnamefont {Kaiser}}, \
  and\ \bibinfo {author} {\bibfnamefont {U.-G.}\ \bibnamefont {Mei{\ss}ner}},\
  }\bibfield  {title} {\enquote {\bibinfo {title} {Aspects of chiral
  pion-nucleon physics},}\ }\href {\doibase 10.1016/S0375-9474(97)00021-3}
  {\bibfield  {journal} {\bibinfo  {journal} {Nucl. Phys. A}\ }\textbf
  {\bibinfo {volume} {615}},\ \bibinfo {pages} {483 -- 500} (\bibinfo {year}
  {1997})}\BibitemShut {NoStop}%
\bibitem [{\citenamefont {Piarulli}\ \emph {et~al.}(2015)\citenamefont
  {Piarulli}, \citenamefont {Girlanda}, \citenamefont {Schiavilla},
  \citenamefont {P\'erez}, \citenamefont {Amaro},\ and\ \citenamefont
  {Arriola}}]{Piarulli2015}%
  \BibitemOpen
  \bibfield  {author} {\bibinfo {author} {\bibfnamefont {M.}~\bibnamefont
  {Piarulli}}, \bibinfo {author} {\bibfnamefont {L.}~\bibnamefont {Girlanda}},
  \bibinfo {author} {\bibfnamefont {R.}~\bibnamefont {Schiavilla}}, \bibinfo
  {author} {\bibfnamefont {R.~Navarro}\ \bibnamefont {P\'erez}}, \bibinfo
  {author} {\bibfnamefont {J.~E.}\ \bibnamefont {Amaro}}, \ and\ \bibinfo
  {author} {\bibfnamefont {E.~Ruiz}\ \bibnamefont {Arriola}},\ }\bibfield
  {title} {\enquote {\bibinfo {title} {Minimally nonlocal nucleon-nucleon
  potentials with chiral two-pion exchange including $\ensuremath{\Delta}$
  resonances},}\ }\href {\doibase 10.1103/PhysRevC.91.024003} {\bibfield
  {journal} {\bibinfo  {journal} {Phys. Rev. C}\ }\textbf {\bibinfo {volume}
  {91}},\ \bibinfo {pages} {024003} (\bibinfo {year} {2015})}\BibitemShut
  {NoStop}%
\bibitem [{\citenamefont {Krebs}\ \emph {et~al.}(2007)\citenamefont {Krebs},
  \citenamefont {Epelbaum},\ and\ \citenamefont {Mei{\ss}ner}}]{Krebs2007}%
  \BibitemOpen
  \bibfield  {author} {\bibinfo {author} {\bibfnamefont {H.}~\bibnamefont
  {Krebs}}, \bibinfo {author} {\bibfnamefont {E.}~\bibnamefont {Epelbaum}}, \
  and\ \bibinfo {author} {\bibfnamefont {U.~G.}\ \bibnamefont {Mei{\ss}ner}},\
  }\bibfield  {title} {\enquote {\bibinfo {title} {Nuclear forces with
  $\mathrm{\ensuremath{\Delta}}$ excitations up to next-to-next-to-leading
  order, part i: Peripheral nucleon-nucleon waves},}\ }\href {\doibase
  10.1140/epja/i2007-10372-y} {\bibfield  {journal} {\bibinfo  {journal} {Eur.
  Phys. J. A}\ }\textbf {\bibinfo {volume} {32}},\ \bibinfo {pages} {127--137}
  (\bibinfo {year} {2007})}\BibitemShut {NoStop}%
\bibitem [{\citenamefont {Piarulli}\ \emph {et~al.}(2016)\citenamefont
  {Piarulli}, \citenamefont {Girlanda}, \citenamefont {Schiavilla},
  \citenamefont {Kievsky}, \citenamefont {Lovato}, \citenamefont {Marcucci},
  \citenamefont {Pieper}, \citenamefont {Viviani},\ and\ \citenamefont
  {Wiringa}}]{Piarulli2016}%
  \BibitemOpen
  \bibfield  {author} {\bibinfo {author} {\bibfnamefont {M.}~\bibnamefont
  {Piarulli}}, \bibinfo {author} {\bibfnamefont {L.}~\bibnamefont {Girlanda}},
  \bibinfo {author} {\bibfnamefont {R.}~\bibnamefont {Schiavilla}}, \bibinfo
  {author} {\bibfnamefont {A.}~\bibnamefont {Kievsky}}, \bibinfo {author}
  {\bibfnamefont {A.}~\bibnamefont {Lovato}}, \bibinfo {author} {\bibfnamefont
  {L.~E.}\ \bibnamefont {Marcucci}}, \bibinfo {author} {\bibfnamefont
  {Steven~C.}\ \bibnamefont {Pieper}}, \bibinfo {author} {\bibfnamefont
  {M.}~\bibnamefont {Viviani}}, \ and\ \bibinfo {author} {\bibfnamefont
  {R.~B.}\ \bibnamefont {Wiringa}},\ }\bibfield  {title} {\enquote {\bibinfo
  {title} {Local chiral potentials with
  $\mathrm{\ensuremath{\Delta}}$-intermediate states and the structure of light
  nuclei},}\ }\href {\doibase 10.1103/PhysRevC.94.054007} {\bibfield  {journal}
  {\bibinfo  {journal} {Phys. Rev. C}\ }\textbf {\bibinfo {volume} {94}},\
  \bibinfo {pages} {054007} (\bibinfo {year} {2016})}\BibitemShut {NoStop}%
\bibitem [{\citenamefont {Logoteta}\ \emph {et~al.}(2016)\citenamefont
  {Logoteta}, \citenamefont {Bombaci},\ and\ \citenamefont
  {Kievsky}}]{Logoteta2016}%
  \BibitemOpen
  \bibfield  {author} {\bibinfo {author} {\bibfnamefont {Domenico}\
  \bibnamefont {Logoteta}}, \bibinfo {author} {\bibfnamefont {Ignazio}\
  \bibnamefont {Bombaci}}, \ and\ \bibinfo {author} {\bibfnamefont {Alejandro}\
  \bibnamefont {Kievsky}},\ }\bibfield  {title} {\enquote {\bibinfo {title}
  {Nuclear matter properties from local chiral interactions with
  $\mathrm{\ensuremath{\Delta}}$ isobar intermediate states},}\ }\href
  {\doibase 10.1103/PhysRevC.94.064001} {\bibfield  {journal} {\bibinfo
  {journal} {Phys. Rev. C}\ }\textbf {\bibinfo {volume} {94}},\ \bibinfo
  {pages} {064001} (\bibinfo {year} {2016})}\BibitemShut {NoStop}%
\bibitem [{\citenamefont {{Piarulli}}\ \emph {et~al.}(2017)\citenamefont
  {{Piarulli}}, \citenamefont {{Baroni}}, \citenamefont {{Girlanda}},
  \citenamefont {{Kievsky}}, \citenamefont {{Lovato}}, \citenamefont {{Lusk}},
  \citenamefont {{Marcucci}}, \citenamefont {{Pieper}}, \citenamefont
  {{Schiavilla}}, \citenamefont {{Viviani}},\ and\ \citenamefont
  {{Wiringa}}}]{piarulli2017}%
  \BibitemOpen
  \bibfield  {author} {\bibinfo {author} {\bibfnamefont {M.}~\bibnamefont
  {{Piarulli}}}, \bibinfo {author} {\bibfnamefont {A.}~\bibnamefont
  {{Baroni}}}, \bibinfo {author} {\bibfnamefont {L.}~\bibnamefont
  {{Girlanda}}}, \bibinfo {author} {\bibfnamefont {A.}~\bibnamefont
  {{Kievsky}}}, \bibinfo {author} {\bibfnamefont {A.}~\bibnamefont {{Lovato}}},
  \bibinfo {author} {\bibfnamefont {E.}~\bibnamefont {{Lusk}}}, \bibinfo
  {author} {\bibfnamefont {L.~E.}\ \bibnamefont {{Marcucci}}}, \bibinfo
  {author} {\bibfnamefont {S.~C.}\ \bibnamefont {{Pieper}}}, \bibinfo {author}
  {\bibfnamefont {R.}~\bibnamefont {{Schiavilla}}}, \bibinfo {author}
  {\bibfnamefont {M.}~\bibnamefont {{Viviani}}}, \ and\ \bibinfo {author}
  {\bibfnamefont {R.~B.}\ \bibnamefont {{Wiringa}}},\ }\bibfield  {title}
  {\enquote {\bibinfo {title} {{Light-nuclei spectra from chiral dynamics}},}\
  }\href {http://adsabs.harvard.edu/abs/2017arXiv170702883P} {\bibfield
  {journal} {\bibinfo  {journal} {ArXiv e-prints}\ } (\bibinfo {year}
  {2017})},\ \Eprint {http://arxiv.org/abs/1707.02883} {arXiv:1707.02883
  [nucl-th]} \BibitemShut {NoStop}%
\bibitem [{\citenamefont {Fujita}\ and\ \citenamefont
  {Miyazawa}(1957)}]{fujita1957}%
  \BibitemOpen
  \bibfield  {author} {\bibinfo {author} {\bibfnamefont {{Jun-ichi}}\
  \bibnamefont {Fujita}}\ and\ \bibinfo {author} {\bibfnamefont {Hironari}\
  \bibnamefont {Miyazawa}},\ }\bibfield  {title} {\enquote {\bibinfo {title}
  {Pion theory of three-body forces},}\ }\href {\doibase 10.1143/PTP.17.360}
  {\bibfield  {journal} {\bibinfo  {journal} {Progress of Theoretical Physics}\
  }\textbf {\bibinfo {volume} {17}},\ \bibinfo {pages} {360--365} (\bibinfo
  {year} {1957})}\BibitemShut {NoStop}%
\bibitem [{\citenamefont {Epelbaum}\ \emph {et~al.}(2008)\citenamefont
  {Epelbaum}, \citenamefont {Krebs},\ and\ \citenamefont
  {Mei{\ss}ner}}]{epelbaum2008}%
  \BibitemOpen
  \bibfield  {author} {\bibinfo {author} {\bibfnamefont {E.}~\bibnamefont
  {Epelbaum}}, \bibinfo {author} {\bibfnamefont {H.}~\bibnamefont {Krebs}}, \
  and\ \bibinfo {author} {\bibfnamefont {U.-G.}\ \bibnamefont {Mei{\ss}ner}},\
  }\bibfield  {title} {\enquote {\bibinfo {title} {{$\Delta$-excitations and
  the three-nucleon force}},}\ }\href {\doibase
  10.1016/j.nuclphysa.2008.02.305} {\bibfield  {journal} {\bibinfo  {journal}
  {Nucl. Phys. A}\ }\textbf {\bibinfo {volume} {806}},\ \bibinfo {pages} {65 --
  78} (\bibinfo {year} {2008})}\BibitemShut {NoStop}%
\bibitem [{\citenamefont {Siemens}\ \emph {et~al.}(2017)\citenamefont
  {Siemens}, \citenamefont {de~Elvira}, \citenamefont {Epelbaum}, \citenamefont
  {Hoferichter}, \citenamefont {Krebs}, \citenamefont {Kubis},\ and\
  \citenamefont {Mei{\ss}ner}}]{Siemens2017}%
  \BibitemOpen
  \bibfield  {author} {\bibinfo {author} {\bibfnamefont {D.}~\bibnamefont
  {Siemens}}, \bibinfo {author} {\bibfnamefont {J.~Ruiz}\ \bibnamefont
  {de~Elvira}}, \bibinfo {author} {\bibfnamefont {E.}~\bibnamefont {Epelbaum}},
  \bibinfo {author} {\bibfnamefont {M.}~\bibnamefont {Hoferichter}}, \bibinfo
  {author} {\bibfnamefont {H.}~\bibnamefont {Krebs}}, \bibinfo {author}
  {\bibfnamefont {B.}~\bibnamefont {Kubis}}, \ and\ \bibinfo {author}
  {\bibfnamefont {U.-G.}\ \bibnamefont {Mei{\ss}ner}},\ }\bibfield  {title}
  {\enquote {\bibinfo {title} {Reconciling threshold and subthreshold
  expansions for pion-nucleon scattering},}\ }\href {\doibase
  10.1016/j.physletb.2017.04.039} {\bibfield  {journal} {\bibinfo  {journal}
  {Phys. Letts. B}\ }\textbf {\bibinfo {volume} {770}},\ \bibinfo {pages} {27
  -- 34} (\bibinfo {year} {2017})}\BibitemShut {NoStop}%
\bibitem [{\citenamefont {Hoferichter}\ \emph {et~al.}(2016)\citenamefont
  {Hoferichter}, \citenamefont {{Ruiz de Elvira}}, \citenamefont {Kubis},\ and\
  \citenamefont {Mei{\ss}ner}}]{hoferichter2016}%
  \BibitemOpen
  \bibfield  {author} {\bibinfo {author} {\bibfnamefont {M.}~\bibnamefont
  {Hoferichter}}, \bibinfo {author} {\bibfnamefont {J.}~\bibnamefont {{Ruiz de
  Elvira}}}, \bibinfo {author} {\bibfnamefont {B.}~\bibnamefont {Kubis}}, \
  and\ \bibinfo {author} {\bibfnamefont {U.-G.}\ \bibnamefont {Mei{\ss}ner}},\
  }\bibfield  {title} {\enquote {\bibinfo {title} {Roy-steiner-equation
  analysis of pion-nucleon scattering},}\ }\href {\doibase
  10.1016/j.physrep.2016.02.002} {\bibfield  {journal} {\bibinfo  {journal}
  {Phys. Rept.}\ }\textbf {\bibinfo {volume} {625}},\ \bibinfo {pages} {1 --
  88} (\bibinfo {year} {2016})}\BibitemShut {NoStop}%
\bibitem [{\citenamefont {Hemmert}\ \emph {et~al.}(1998)\citenamefont
  {Hemmert}, \citenamefont {Holstein},\ and\ \citenamefont
  {Kambor}}]{Hemmert1998}%
  \BibitemOpen
  \bibfield  {author} {\bibinfo {author} {\bibfnamefont {T.~R.}\ \bibnamefont
  {Hemmert}}, \bibinfo {author} {\bibfnamefont {B.~R.}\ \bibnamefont
  {Holstein}}, \ and\ \bibinfo {author} {\bibfnamefont {J.}~\bibnamefont
  {Kambor}},\ }\bibfield  {title} {\enquote {\bibinfo {title} {Heavy baryon
  chiral perturbation theory with light deltas},}\ }\href
  {http://stacks.iop.org/0954-3899/24/i=10/a=003} {\bibfield  {journal}
  {\bibinfo  {journal} {J. Phys. G: Nucl. Part. Phys.}\ }\textbf {\bibinfo
  {volume} {24}},\ \bibinfo {pages} {1831} (\bibinfo {year}
  {1998})}\BibitemShut {NoStop}%
\bibitem [{\citenamefont {Kaiser}\ \emph {et~al.}(1998)\citenamefont {Kaiser},
  \citenamefont {Gerstend{\"o}rfer},\ and\ \citenamefont {Weise}}]{Kaiser1998}%
  \BibitemOpen
  \bibfield  {author} {\bibinfo {author} {\bibfnamefont {N.}~\bibnamefont
  {Kaiser}}, \bibinfo {author} {\bibfnamefont {S.}~\bibnamefont
  {Gerstend{\"o}rfer}}, \ and\ \bibinfo {author} {\bibfnamefont
  {W.}~\bibnamefont {Weise}},\ }\bibfield  {title} {\enquote {\bibinfo {title}
  {Peripheral nn-scattering: role of delta-excitation, correlated two-pion and
  vector meson exchange},}\ }\href {\doibase 10.1016/S0375-9474(98)00234-6}
  {\bibfield  {journal} {\bibinfo  {journal} {Nucl. Phys. A}\ }\textbf
  {\bibinfo {volume} {637}},\ \bibinfo {pages} {395 -- 420} (\bibinfo {year}
  {1998})}\BibitemShut {NoStop}%
\bibitem [{\citenamefont {Long}\ and\ \citenamefont {Lensky}(2011)}]{Long2011}%
  \BibitemOpen
  \bibfield  {author} {\bibinfo {author} {\bibfnamefont {Bingwei}\ \bibnamefont
  {Long}}\ and\ \bibinfo {author} {\bibfnamefont {Vadim}\ \bibnamefont
  {Lensky}},\ }\bibfield  {title} {\enquote {\bibinfo {title} {Heavy-particle
  formalism with foldy-wouthuysen representation},}\ }\href {\doibase
  10.1103/PhysRevC.83.045206} {\bibfield  {journal} {\bibinfo  {journal} {Phys.
  Rev. C}\ }\textbf {\bibinfo {volume} {83}},\ \bibinfo {pages} {045206}
  (\bibinfo {year} {2011})}\BibitemShut {NoStop}%
\bibitem [{\citenamefont {Pandharipande}\ \emph {et~al.}(2005)\citenamefont
  {Pandharipande}, \citenamefont {Phillips},\ and\ \citenamefont
  {Kolck}}]{pandharipande2005}%
  \BibitemOpen
  \bibfield  {author} {\bibinfo {author} {\bibfnamefont {V.~R.}\ \bibnamefont
  {Pandharipande}}, \bibinfo {author} {\bibfnamefont {D.~R.}\ \bibnamefont
  {Phillips}}, \ and\ \bibinfo {author} {\bibfnamefont {U.~van}\ \bibnamefont
  {Kolck}},\ }\bibfield  {title} {\enquote {\bibinfo {title}
  {\ensuremath{\Delta} effects in pion-nucleon scattering and the strength of
  the two-pion-exchange three-nucleon interaction},}\ }\href {\doibase
  10.1103/PhysRevC.71.064002} {\bibfield  {journal} {\bibinfo  {journal} {Phys.
  Rev. C}\ }\textbf {\bibinfo {volume} {71}},\ \bibinfo {pages} {064002}
  (\bibinfo {year} {2005})}\BibitemShut {NoStop}%
\bibitem [{\citenamefont {P\'erez}\ \emph {et~al.}(2013)\citenamefont
  {P\'erez}, \citenamefont {Amaro},\ and\ \citenamefont
  {Arriola}}]{Navarro2013}%
  \BibitemOpen
  \bibfield  {author} {\bibinfo {author} {\bibfnamefont {R.~Navarro}\
  \bibnamefont {P\'erez}}, \bibinfo {author} {\bibfnamefont {J.~E.}\
  \bibnamefont {Amaro}}, \ and\ \bibinfo {author} {\bibfnamefont {E.~Ruiz}\
  \bibnamefont {Arriola}},\ }\bibfield  {title} {\enquote {\bibinfo {title}
  {Coarse-grained potential analysis of neutron-proton and proton-proton
  scattering below the pion production threshold},}\ }\href {\doibase
  10.1103/PhysRevC.88.064002} {\bibfield  {journal} {\bibinfo  {journal} {Phys.
  Rev. C}\ }\textbf {\bibinfo {volume} {88}},\ \bibinfo {pages} {064002}
  (\bibinfo {year} {2013})}\BibitemShut {NoStop}%
\bibitem [{\citenamefont {Epelbaum}\ \emph {et~al.}(2005)\citenamefont
  {Epelbaum}, \citenamefont {Glöckle},\ and\ \citenamefont
  {Meißner}}]{epelbaum2005}%
  \BibitemOpen
  \bibfield  {author} {\bibinfo {author} {\bibfnamefont {E.}~\bibnamefont
  {Epelbaum}}, \bibinfo {author} {\bibfnamefont {W.}~\bibnamefont {Glöckle}},
  \ and\ \bibinfo {author} {\bibfnamefont {Ulf-G.}\ \bibnamefont {Meißner}},\
  }\bibfield  {title} {\enquote {\bibinfo {title} {The two-nucleon system at
  next-to-next-to-next-to-leading order},}\ }\href {\doibase
  http://dx.doi.org/10.1016/j.nuclphysa.2004.09.107} {\bibfield  {journal}
  {\bibinfo  {journal} {Nuclear Physics A}\ }\textbf {\bibinfo {volume}
  {747}},\ \bibinfo {pages} {362 -- 424} (\bibinfo {year} {2005})}\BibitemShut
  {NoStop}%
\bibitem [{\citenamefont {Valderrama}\ and\ \citenamefont
  {Arriola}(2009)}]{valderrama2009}%
  \BibitemOpen
  \bibfield  {author} {\bibinfo {author} {\bibfnamefont {M.~Pav\'on}\
  \bibnamefont {Valderrama}}\ and\ \bibinfo {author} {\bibfnamefont {E.~Ruiz}\
  \bibnamefont {Arriola}},\ }\bibfield  {title} {\enquote {\bibinfo {title}
  {Renormalization of chiral two-pion exchange $\mathit{NN}$ interactions with
  \ensuremath{\Delta} excitations: Central phases and the deuteron},}\ }\href
  {\doibase 10.1103/PhysRevC.79.044001} {\bibfield  {journal} {\bibinfo
  {journal} {Phys. Rev. C}\ }\textbf {\bibinfo {volume} {79}},\ \bibinfo
  {pages} {044001} (\bibinfo {year} {2009})}\BibitemShut {NoStop}%
\bibitem [{\citenamefont {Pavon~Valderrama}\ and\ \citenamefont
  {Ruiz~Arriola}(2011)}]{valderrama2011}%
  \BibitemOpen
  \bibfield  {author} {\bibinfo {author} {\bibfnamefont {M.}~\bibnamefont
  {Pavon~Valderrama}}\ and\ \bibinfo {author} {\bibfnamefont {E.}~\bibnamefont
  {Ruiz~Arriola}},\ }\bibfield  {title} {\enquote {\bibinfo {title}
  {Renormalization of chiral two-pion exchange $\mathit{NN}$ interactions with
  $\ensuremath{\Delta}$ excitations: Correlations in the partial-wave
  expansion},}\ }\href {\doibase 10.1103/PhysRevC.83.044002} {\bibfield
  {journal} {\bibinfo  {journal} {Phys. Rev. C}\ }\textbf {\bibinfo {volume}
  {83}},\ \bibinfo {pages} {044002} (\bibinfo {year} {2011})}\BibitemShut
  {NoStop}%
\bibitem [{\citenamefont {Epelbaum}\ \emph
  {et~al.}(2015{\natexlab{b}})\citenamefont {Epelbaum}, \citenamefont {Krebs},\
  and\ \citenamefont {Mei{\ss}ner}}]{epelbaum2015b}%
  \BibitemOpen
  \bibfield  {author} {\bibinfo {author} {\bibfnamefont {E.}~\bibnamefont
  {Epelbaum}}, \bibinfo {author} {\bibfnamefont {H.}~\bibnamefont {Krebs}}, \
  and\ \bibinfo {author} {\bibfnamefont {U.~G.}\ \bibnamefont {Mei{\ss}ner}},\
  }\bibfield  {title} {\enquote {\bibinfo {title} {Improved chiral
  nucleon-nucleon potential up to next-to-next-to-next-to-leading order},}\
  }\href {\doibase 10.1140/epja/i2015-15053-8} {\bibfield  {journal} {\bibinfo
  {journal} {The European Physical Journal A}\ }\textbf {\bibinfo {volume}
  {51}},\ \bibinfo {pages} {53} (\bibinfo {year}
  {2015}{\natexlab{b}})}\BibitemShut {NoStop}%
\bibitem [{\citenamefont {Dyhdalo}\ \emph {et~al.}(2016)\citenamefont
  {Dyhdalo}, \citenamefont {Furnstahl}, \citenamefont {Hebeler},\ and\
  \citenamefont {Tews}}]{dyhaldo2016}%
  \BibitemOpen
  \bibfield  {author} {\bibinfo {author} {\bibfnamefont {A.}~\bibnamefont
  {Dyhdalo}}, \bibinfo {author} {\bibfnamefont {R.~J.}\ \bibnamefont
  {Furnstahl}}, \bibinfo {author} {\bibfnamefont {K.}~\bibnamefont {Hebeler}},
  \ and\ \bibinfo {author} {\bibfnamefont {I.}~\bibnamefont {Tews}},\
  }\bibfield  {title} {\enquote {\bibinfo {title} {Regulator artifacts in
  uniform matter for chiral interactions},}\ }\href {\doibase
  10.1103/PhysRevC.94.034001} {\bibfield  {journal} {\bibinfo  {journal} {Phys.
  Rev. C}\ }\textbf {\bibinfo {volume} {94}},\ \bibinfo {pages} {034001}
  (\bibinfo {year} {2016})}\BibitemShut {NoStop}%
\bibitem [{\citenamefont {Hoppe}\ \emph {et~al.}(2017)\citenamefont {Hoppe},
  \citenamefont {Drischler}, \citenamefont {Furnstahl}, \citenamefont
  {Hebeler},\ and\ \citenamefont {Schwenk}}]{hoppe2017}%
  \BibitemOpen
  \bibfield  {author} {\bibinfo {author} {\bibfnamefont {J.}~\bibnamefont
  {Hoppe}}, \bibinfo {author} {\bibfnamefont {C.}~\bibnamefont {Drischler}},
  \bibinfo {author} {\bibfnamefont {R.~J.}\ \bibnamefont {Furnstahl}}, \bibinfo
  {author} {\bibfnamefont {K.}~\bibnamefont {Hebeler}}, \ and\ \bibinfo
  {author} {\bibfnamefont {A.}~\bibnamefont {Schwenk}},\ }\bibfield  {title}
  {\enquote {\bibinfo {title} {Weinberg eigenvalues for chiral nucleon-nucleon
  interactions},}\ }\href {\doibase 10.1103/PhysRevC.96.054002} {\bibfield
  {journal} {\bibinfo  {journal} {Phys. Rev. C}\ }\textbf {\bibinfo {volume}
  {96}},\ \bibinfo {pages} {054002} (\bibinfo {year} {2017})}\BibitemShut
  {NoStop}%
\bibitem [{\citenamefont {Ekstr{\"o}m}\ \emph {et~al.}(2015)\citenamefont
  {Ekstr{\"o}m}, \citenamefont {Carlsson}, \citenamefont {Wendt}, \citenamefont
  {Forss{\'e}n}, \citenamefont {Jensen}, \citenamefont {Machleidt},\ and\
  \citenamefont {Wild}}]{ekstrom2015b}%
  \BibitemOpen
  \bibfield  {author} {\bibinfo {author} {\bibfnamefont {A.}~\bibnamefont
  {Ekstr{\"o}m}}, \bibinfo {author} {\bibfnamefont {B.~D.}\ \bibnamefont
  {Carlsson}}, \bibinfo {author} {\bibfnamefont {K.~A.}\ \bibnamefont {Wendt}},
  \bibinfo {author} {\bibfnamefont {C.}~\bibnamefont {Forss{\'e}n}}, \bibinfo
  {author} {\bibfnamefont {M.~Hjorth}\ \bibnamefont {Jensen}}, \bibinfo
  {author} {\bibfnamefont {R.}~\bibnamefont {Machleidt}}, \ and\ \bibinfo
  {author} {\bibfnamefont {S.~M.}\ \bibnamefont {Wild}},\ }\bibfield  {title}
  {\enquote {\bibinfo {title} {Statistical uncertainties of a chiral
  interaction at next-to-next-to leading order},}\ }\href {\doibase
  10.1088/0954-3899/42/3/034003} {\bibfield  {journal} {\bibinfo  {journal} {J.
  Phys. G}\ }\textbf {\bibinfo {volume} {42}},\ \bibinfo {pages} {034003}
  (\bibinfo {year} {2015})}\BibitemShut {NoStop}%
\bibitem [{\citenamefont {P\'erez}\ \emph {et~al.}(2015)\citenamefont
  {P\'erez}, \citenamefont {Amaro},\ and\ \citenamefont
  {Arriola}}]{Navarro2015}%
  \BibitemOpen
  \bibfield  {author} {\bibinfo {author} {\bibfnamefont {R.~Navarro}\
  \bibnamefont {P\'erez}}, \bibinfo {author} {\bibfnamefont {J.~E.}\
  \bibnamefont {Amaro}}, \ and\ \bibinfo {author} {\bibfnamefont {E.~Ruiz}\
  \bibnamefont {Arriola}},\ }\bibfield  {title} {\enquote {\bibinfo {title}
  {Low-energy chiral two-pion exchange potential with statistical
  uncertainties},}\ }\href {\doibase 10.1103/PhysRevC.91.054002} {\bibfield
  {journal} {\bibinfo  {journal} {Phys. Rev. C}\ }\textbf {\bibinfo {volume}
  {91}},\ \bibinfo {pages} {054002} (\bibinfo {year} {2015})}\BibitemShut
  {NoStop}%
\bibitem [{\citenamefont {Furnstahl}\ \emph {et~al.}(2015)\citenamefont
  {Furnstahl}, \citenamefont {Klco}, \citenamefont {Phillips},\ and\
  \citenamefont {Wesolowski}}]{furnstahl2015}%
  \BibitemOpen
  \bibfield  {author} {\bibinfo {author} {\bibfnamefont {R.~J.}\ \bibnamefont
  {Furnstahl}}, \bibinfo {author} {\bibfnamefont {N.}~\bibnamefont {Klco}},
  \bibinfo {author} {\bibfnamefont {D.~R.}\ \bibnamefont {Phillips}}, \ and\
  \bibinfo {author} {\bibfnamefont {S.}~\bibnamefont {Wesolowski}},\ }\bibfield
   {title} {\enquote {\bibinfo {title} {Quantifying truncation errors in
  effective field theory},}\ }\href {\doibase 10.1103/PhysRevC.92.024005}
  {\bibfield  {journal} {\bibinfo  {journal} {Phys. Rev. C}\ }\textbf {\bibinfo
  {volume} {92}},\ \bibinfo {pages} {024005} (\bibinfo {year}
  {2015})}\BibitemShut {NoStop}%
\bibitem [{\citenamefont {Navr\'atil}\ \emph {et~al.}(2000)\citenamefont
  {Navr\'atil}, \citenamefont {Kamuntavi\ifmmode~\check{c}\else
  \v{c}\fi{}ius},\ and\ \citenamefont {Barrett}}]{kamuntavicius2000}%
  \BibitemOpen
  \bibfield  {author} {\bibinfo {author} {\bibfnamefont {P.}~\bibnamefont
  {Navr\'atil}}, \bibinfo {author} {\bibfnamefont {G.~P.}\ \bibnamefont
  {Kamuntavi\ifmmode~\check{c}\else \v{c}\fi{}ius}}, \ and\ \bibinfo {author}
  {\bibfnamefont {B.~R.}\ \bibnamefont {Barrett}},\ }\bibfield  {title}
  {\enquote {\bibinfo {title} {Few-nucleon systems in a translationally
  invariant harmonic oscillator basis},}\ }\href {\doibase
  10.1103/PhysRevC.61.044001} {\bibfield  {journal} {\bibinfo  {journal} {Phys.
  Rev. C}\ }\textbf {\bibinfo {volume} {61}},\ \bibinfo {pages} {044001}
  (\bibinfo {year} {2000})}\BibitemShut {NoStop}%
\bibitem [{\citenamefont {Angeli}\ and\ \citenamefont
  {Marinova}(2013)}]{angeli2013}%
  \BibitemOpen
  \bibfield  {author} {\bibinfo {author} {\bibfnamefont {I.}~\bibnamefont
  {Angeli}}\ and\ \bibinfo {author} {\bibfnamefont {K.P.}\ \bibnamefont
  {Marinova}},\ }\bibfield  {title} {\enquote {\bibinfo {title} {Table of
  experimental nuclear ground state charge radii: An update},}\ }\href
  {\doibase 10.1016/j.adt.2011.12.006} {\bibfield  {journal} {\bibinfo
  {journal} {At. Data Nucl. Data Tables}\ }\textbf {\bibinfo {volume} {99}},\
  \bibinfo {pages} {69 -- 95} (\bibinfo {year} {2013})}\BibitemShut {NoStop}%
\bibitem [{\citenamefont {K{\"u}mmel}\ \emph {et~al.}(1978)\citenamefont
  {K{\"u}mmel}, \citenamefont {L{\"u}hrmann},\ and\ \citenamefont
  {Zabolitzky}}]{kuemmel1978}%
  \BibitemOpen
  \bibfield  {author} {\bibinfo {author} {\bibfnamefont {H.}~\bibnamefont
  {K{\"u}mmel}}, \bibinfo {author} {\bibfnamefont {K.~H.}\ \bibnamefont
  {L{\"u}hrmann}}, \ and\ \bibinfo {author} {\bibfnamefont {J.~G.}\
  \bibnamefont {Zabolitzky}},\ }\bibfield  {title} {\enquote {\bibinfo {title}
  {{Many-fermion theory in $\exp S$ (or coupled cluster) form}},}\ }\href
  {\doibase 10.1016/0370-1573(78)90081-9} {\bibfield  {journal} {\bibinfo
  {journal} {Phys. Rep.}\ }\textbf {\bibinfo {volume} {36}},\ \bibinfo {pages}
  {1 -- 63} (\bibinfo {year} {1978})}\BibitemShut {NoStop}%
\bibitem [{\citenamefont {Bishop}(1991)}]{bishop1991}%
  \BibitemOpen
  \bibfield  {author} {\bibinfo {author} {\bibfnamefont {R.~F.}\ \bibnamefont
  {Bishop}},\ }\bibfield  {title} {\enquote {\bibinfo {title} {An overview of
  coupled cluster theory and its applications in physics},}\ }\href {\doibase
  10.1007/BF01119617} {\bibfield  {journal} {\bibinfo  {journal} {Theor. Chim.
  Acta}\ }\textbf {\bibinfo {volume} {80}},\ \bibinfo {pages} {95--148}
  (\bibinfo {year} {1991})}\BibitemShut {NoStop}%
\bibitem [{\citenamefont {Bartlett}\ and\ \citenamefont
  {Musia\l{}}(2007)}]{bartlett2007}%
  \BibitemOpen
  \bibfield  {author} {\bibinfo {author} {\bibfnamefont {R.~J.}\ \bibnamefont
  {Bartlett}}\ and\ \bibinfo {author} {\bibfnamefont {M.}~\bibnamefont
  {Musia\l{}}},\ }\bibfield  {title} {\enquote {\bibinfo {title}
  {Coupled-cluster theory in quantum chemistry},}\ }\href {\doibase
  10.1103/RevModPhys.79.291} {\bibfield  {journal} {\bibinfo  {journal} {Rev.
  Mod. Phys.}\ }\textbf {\bibinfo {volume} {79}},\ \bibinfo {pages} {291--352}
  (\bibinfo {year} {2007})}\BibitemShut {NoStop}%
\bibitem [{\citenamefont {Hagen}\ \emph {et~al.}(2009)\citenamefont {Hagen},
  \citenamefont {Papenbrock},\ and\ \citenamefont {Dean}}]{hagen2009a}%
  \BibitemOpen
  \bibfield  {author} {\bibinfo {author} {\bibfnamefont {G.}~\bibnamefont
  {Hagen}}, \bibinfo {author} {\bibfnamefont {T.}~\bibnamefont {Papenbrock}}, \
  and\ \bibinfo {author} {\bibfnamefont {D.~J.}\ \bibnamefont {Dean}},\
  }\bibfield  {title} {\enquote {\bibinfo {title} {Solution of the
  center-of-mass problem in nuclear structure calculations},}\ }\href {\doibase
  10.1103/PhysRevLett.103.062503} {\bibfield  {journal} {\bibinfo  {journal}
  {Phys. Rev. Lett.}\ }\textbf {\bibinfo {volume} {103}},\ \bibinfo {pages}
  {062503} (\bibinfo {year} {2009})}\BibitemShut {NoStop}%
\bibitem [{\citenamefont {Hagen}\ \emph {et~al.}(2010)\citenamefont {Hagen},
  \citenamefont {Papenbrock}, \citenamefont {Dean},\ and\ \citenamefont
  {Hjorth-Jensen}}]{hagen2010b}%
  \BibitemOpen
  \bibfield  {author} {\bibinfo {author} {\bibfnamefont {G.}~\bibnamefont
  {Hagen}}, \bibinfo {author} {\bibfnamefont {T.}~\bibnamefont {Papenbrock}},
  \bibinfo {author} {\bibfnamefont {D.~J.}\ \bibnamefont {Dean}}, \ and\
  \bibinfo {author} {\bibfnamefont {M.}~\bibnamefont {Hjorth-Jensen}},\
  }\bibfield  {title} {\enquote {\bibinfo {title} {\textit{Ab initio}
  coupled-cluster approach to nuclear structure with modern nucleon-nucleon
  interactions},}\ }\href {\doibase 10.1103/PhysRevC.82.034330} {\bibfield
  {journal} {\bibinfo  {journal} {Phys. Rev. C}\ }\textbf {\bibinfo {volume}
  {82}},\ \bibinfo {pages} {034330} (\bibinfo {year} {2010})}\BibitemShut
  {NoStop}%
\bibitem [{\citenamefont {Jansen}(2013)}]{jansen2012}%
  \BibitemOpen
  \bibfield  {author} {\bibinfo {author} {\bibfnamefont {G.~R.}\ \bibnamefont
  {Jansen}},\ }\bibfield  {title} {\enquote {\bibinfo {title} {Spherical
  coupled-cluster theory for open-shell nuclei},}\ }\href {\doibase
  10.1103/PhysRevC.88.024305} {\bibfield  {journal} {\bibinfo  {journal} {Phys.
  Rev. C}\ }\textbf {\bibinfo {volume} {88}},\ \bibinfo {pages} {024305}
  (\bibinfo {year} {2013})}\BibitemShut {NoStop}%
\bibitem [{\citenamefont {Morris}\ \emph {et~al.}(2015)\citenamefont {Morris},
  \citenamefont {Parzuchowski},\ and\ \citenamefont {Bogner}}]{morris2015}%
  \BibitemOpen
  \bibfield  {author} {\bibinfo {author} {\bibfnamefont {T.~D.}\ \bibnamefont
  {Morris}}, \bibinfo {author} {\bibfnamefont {N.~M.}\ \bibnamefont
  {Parzuchowski}}, \ and\ \bibinfo {author} {\bibfnamefont {S.~K.}\
  \bibnamefont {Bogner}},\ }\bibfield  {title} {\enquote {\bibinfo {title}
  {Magnus expansion and in-medium similarity renormalization group},}\ }\href
  {\doibase 10.1103/PhysRevC.92.034331} {\bibfield  {journal} {\bibinfo
  {journal} {Phys. Rev. C}\ }\textbf {\bibinfo {volume} {92}},\ \bibinfo
  {pages} {034331} (\bibinfo {year} {2015})}\BibitemShut {NoStop}%
\bibitem [{\citenamefont {Taube}\ and\ \citenamefont
  {Bartlett}(2008)}]{taube2008}%
  \BibitemOpen
  \bibfield  {author} {\bibinfo {author} {\bibfnamefont {A.~G.}\ \bibnamefont
  {Taube}}\ and\ \bibinfo {author} {\bibfnamefont {R.~J.}\ \bibnamefont
  {Bartlett}},\ }\bibfield  {title} {\enquote {\bibinfo {title} {Improving upon
  ccsd(t): Lambda ccsd(t). i. potential energy surfaces},}\ }\href {\doibase
  10.1063/1.2830236} {\bibfield  {journal} {\bibinfo  {journal} {J. Chem.
  Phys.}\ }\textbf {\bibinfo {volume} {128}},\ \bibinfo {eid} {044110}
  (\bibinfo {year} {2008})}\BibitemShut {NoStop}%
\bibitem [{\citenamefont {Binder}\ \emph {et~al.}(2013)\citenamefont {Binder},
  \citenamefont {Piecuch}, \citenamefont {Calci}, \citenamefont {Langhammer},
  \citenamefont {Navr\'atil},\ and\ \citenamefont {Roth}}]{binder2013}%
  \BibitemOpen
  \bibfield  {author} {\bibinfo {author} {\bibfnamefont {S.}~\bibnamefont
  {Binder}}, \bibinfo {author} {\bibfnamefont {P.}~\bibnamefont {Piecuch}},
  \bibinfo {author} {\bibfnamefont {A.}~\bibnamefont {Calci}}, \bibinfo
  {author} {\bibfnamefont {J.}~\bibnamefont {Langhammer}}, \bibinfo {author}
  {\bibfnamefont {P.}~\bibnamefont {Navr\'atil}}, \ and\ \bibinfo {author}
  {\bibfnamefont {R.}~\bibnamefont {Roth}},\ }\bibfield  {title} {\enquote
  {\bibinfo {title} {Extension of coupled-cluster theory with a noniterative
  treatment of connected triply excited clusters to three-body hamiltonians},}\
  }\href {\doibase 10.1103/PhysRevC.88.054319} {\bibfield  {journal} {\bibinfo
  {journal} {Phys. Rev. C}\ }\textbf {\bibinfo {volume} {88}},\ \bibinfo
  {pages} {054319} (\bibinfo {year} {2013})}\BibitemShut {NoStop}%
\bibitem [{\citenamefont {Hagen}\ \emph {et~al.}(2007)\citenamefont {Hagen},
  \citenamefont {Papenbrock}, \citenamefont {Dean}, \citenamefont {Schwenk},
  \citenamefont {Nogga}, \citenamefont {W\l{}och},\ and\ \citenamefont
  {Piecuch}}]{hagen2007a}%
  \BibitemOpen
  \bibfield  {author} {\bibinfo {author} {\bibfnamefont {G.}~\bibnamefont
  {Hagen}}, \bibinfo {author} {\bibfnamefont {T.}~\bibnamefont {Papenbrock}},
  \bibinfo {author} {\bibfnamefont {D.~J.}\ \bibnamefont {Dean}}, \bibinfo
  {author} {\bibfnamefont {A.}~\bibnamefont {Schwenk}}, \bibinfo {author}
  {\bibfnamefont {A.}~\bibnamefont {Nogga}}, \bibinfo {author} {\bibfnamefont
  {M.}~\bibnamefont {W\l{}och}}, \ and\ \bibinfo {author} {\bibfnamefont
  {P.}~\bibnamefont {Piecuch}},\ }\bibfield  {title} {\enquote {\bibinfo
  {title} {{Coupled-cluster theory for three-body Hamiltonians}},}\ }\href
  {\doibase 10.1103/PhysRevC.76.034302} {\bibfield  {journal} {\bibinfo
  {journal} {Phys. Rev. C}\ }\textbf {\bibinfo {volume} {76}},\ \bibinfo
  {pages} {034302} (\bibinfo {year} {2007})}\BibitemShut {NoStop}%
\bibitem [{\citenamefont {Roth}\ \emph {et~al.}(2012)\citenamefont {Roth},
  \citenamefont {Binder}, \citenamefont {Vobig}, \citenamefont {Calci},
  \citenamefont {Langhammer},\ and\ \citenamefont {Navr\'atil}}]{roth2012}%
  \BibitemOpen
  \bibfield  {author} {\bibinfo {author} {\bibfnamefont {R.}~\bibnamefont
  {Roth}}, \bibinfo {author} {\bibfnamefont {S.}~\bibnamefont {Binder}},
  \bibinfo {author} {\bibfnamefont {K.}~\bibnamefont {Vobig}}, \bibinfo
  {author} {\bibfnamefont {A.}~\bibnamefont {Calci}}, \bibinfo {author}
  {\bibfnamefont {J.}~\bibnamefont {Langhammer}}, \ and\ \bibinfo {author}
  {\bibfnamefont {P.}~\bibnamefont {Navr\'atil}},\ }\bibfield  {title}
  {\enquote {\bibinfo {title} {{Medium-Mass Nuclei with Normal-Ordered Chiral
  $NN\mathbf{+}3N$ Interactions}},}\ }\href {\doibase
  10.1103/PhysRevLett.109.052501} {\bibfield  {journal} {\bibinfo  {journal}
  {Phys. Rev. Lett.}\ }\textbf {\bibinfo {volume} {109}},\ \bibinfo {pages}
  {052501} (\bibinfo {year} {2012})}\BibitemShut {NoStop}%
\bibitem [{\citenamefont {Binder}\ \emph {et~al.}(2016)\citenamefont {Binder},
  \citenamefont {Calci}, \citenamefont {Epelbaum}, \citenamefont {Furnstahl},
  \citenamefont {Golak}, \citenamefont {Hebeler}, \citenamefont {Kamada},
  \citenamefont {Krebs}, \citenamefont {Langhammer}, \citenamefont {Liebig},
  \citenamefont {Maris}, \citenamefont {Mei\ss{}ner}, \citenamefont {Minossi},
  \citenamefont {Nogga}, \citenamefont {Potter}, \citenamefont {Roth},
  \citenamefont {Skibi\ifmmode~\acute{n}\else \'{n}\fi{}ski}, \citenamefont
  {Topolnicki}, \citenamefont {Vary},\ and\ \citenamefont
  {Wita\l{}a}}]{binder2015}%
  \BibitemOpen
  \bibfield  {author} {\bibinfo {author} {\bibfnamefont {S.}~\bibnamefont
  {Binder}}, \bibinfo {author} {\bibfnamefont {A.}~\bibnamefont {Calci}},
  \bibinfo {author} {\bibfnamefont {E.}~\bibnamefont {Epelbaum}}, \bibinfo
  {author} {\bibfnamefont {R.~J.}\ \bibnamefont {Furnstahl}}, \bibinfo {author}
  {\bibfnamefont {J.}~\bibnamefont {Golak}}, \bibinfo {author} {\bibfnamefont
  {K.}~\bibnamefont {Hebeler}}, \bibinfo {author} {\bibfnamefont
  {H.}~\bibnamefont {Kamada}}, \bibinfo {author} {\bibfnamefont
  {H.}~\bibnamefont {Krebs}}, \bibinfo {author} {\bibfnamefont
  {J.}~\bibnamefont {Langhammer}}, \bibinfo {author} {\bibfnamefont
  {S.}~\bibnamefont {Liebig}}, \bibinfo {author} {\bibfnamefont
  {P.}~\bibnamefont {Maris}}, \bibinfo {author} {\bibfnamefont {Ulf-G.}\
  \bibnamefont {Mei\ss{}ner}}, \bibinfo {author} {\bibfnamefont
  {D.}~\bibnamefont {Minossi}}, \bibinfo {author} {\bibfnamefont
  {A.}~\bibnamefont {Nogga}}, \bibinfo {author} {\bibfnamefont
  {H.}~\bibnamefont {Potter}}, \bibinfo {author} {\bibfnamefont
  {R.}~\bibnamefont {Roth}}, \bibinfo {author} {\bibfnamefont {R.}~\bibnamefont
  {Skibi\ifmmode~\acute{n}\else \'{n}\fi{}ski}}, \bibinfo {author}
  {\bibfnamefont {K.}~\bibnamefont {Topolnicki}}, \bibinfo {author}
  {\bibfnamefont {J.~P.}\ \bibnamefont {Vary}}, \ and\ \bibinfo {author}
  {\bibfnamefont {H.}~\bibnamefont {Wita\l{}a}} (\bibinfo {collaboration}
  {LENPIC Collaboration}),\ }\bibfield  {title} {\enquote {\bibinfo {title}
  {Few-nucleon systems with state-of-the-art chiral nucleon-nucleon forces},}\
  }\href {\doibase 10.1103/PhysRevC.93.044002} {\bibfield  {journal} {\bibinfo
  {journal} {Phys. Rev. C}\ }\textbf {\bibinfo {volume} {93}},\ \bibinfo
  {pages} {044002} (\bibinfo {year} {2016})}\BibitemShut {NoStop}%
\bibitem [{\citenamefont {Stetcu}\ \emph {et~al.}(2007)\citenamefont {Stetcu},
  \citenamefont {Barrett},\ and\ \citenamefont {van Kolck}}]{stetcu2007}%
  \BibitemOpen
  \bibfield  {author} {\bibinfo {author} {\bibfnamefont {I.}~\bibnamefont
  {Stetcu}}, \bibinfo {author} {\bibfnamefont {B.~R.}\ \bibnamefont {Barrett}},
  \ and\ \bibinfo {author} {\bibfnamefont {U.}~\bibnamefont {van Kolck}},\
  }\bibfield  {title} {\enquote {\bibinfo {title} {No-core shell model in an
  effective-field-theory framework},}\ }\href {\doibase
  10.1016/j.physletb.2007.07.065} {\bibfield  {journal} {\bibinfo  {journal}
  {Phys. Lett. B}\ }\textbf {\bibinfo {volume} {653}},\ \bibinfo {pages} {358
  -- 362} (\bibinfo {year} {2007})}\BibitemShut {NoStop}%
\bibitem [{\citenamefont {Contessi}\ \emph {et~al.}(2017)\citenamefont
  {Contessi}, \citenamefont {Lovato}, \citenamefont {Pederiva}, \citenamefont
  {Roggero}, \citenamefont {Kirscher},\ and\ \citenamefont {van
  Kolck}}]{contessi2017}%
  \BibitemOpen
  \bibfield  {author} {\bibinfo {author} {\bibfnamefont {L.}~\bibnamefont
  {Contessi}}, \bibinfo {author} {\bibfnamefont {A.}~\bibnamefont {Lovato}},
  \bibinfo {author} {\bibfnamefont {F.}~\bibnamefont {Pederiva}}, \bibinfo
  {author} {\bibfnamefont {A.}~\bibnamefont {Roggero}}, \bibinfo {author}
  {\bibfnamefont {J.}~\bibnamefont {Kirscher}}, \ and\ \bibinfo {author}
  {\bibfnamefont {U.}~\bibnamefont {van Kolck}},\ }\bibfield  {title} {\enquote
  {\bibinfo {title} {Ground-state properties of 4he and 16o extrapolated from
  lattice qcd with pionless eft},}\ }\href {\doibase
  https://doi.org/10.1016/j.physletb.2017.07.048} {\bibfield  {journal}
  {\bibinfo  {journal} {Physics Letters B}\ }\textbf {\bibinfo {volume}
  {772}},\ \bibinfo {pages} {839 -- 848} (\bibinfo {year} {2017})}\BibitemShut
  {NoStop}%
\bibitem [{\citenamefont {{Bansal}}\ \emph {et~al.}(2017)\citenamefont
  {{Bansal}}, \citenamefont {{Ekstr{\"o}m}}, \citenamefont {{Hagen}},
  \citenamefont {{Jansen}},\ and\ \citenamefont {{Papenbrock}}}]{bansal2017}%
  \BibitemOpen
  \bibfield  {author} {\bibinfo {author} {\bibfnamefont {S.}~\bibnamefont
  {{Bansal}}, \bibfnamefont {A.~{Binder}}}, \bibinfo {author} {\bibfnamefont
  {A.}~\bibnamefont {{Ekstr{\"o}m}}}, \bibinfo {author} {\bibfnamefont
  {G.}~\bibnamefont {{Hagen}}}, \bibinfo {author} {\bibfnamefont {G.~R.}\
  \bibnamefont {{Jansen}}}, \ and\ \bibinfo {author} {\bibfnamefont
  {T.}~\bibnamefont {{Papenbrock}}},\ }\bibfield  {title} {\enquote {\bibinfo
  {title} {{Pion-less effective field theory for atomic nuclei and lattice
  nuclei}},}\ }\href {http://adsabs.harvard.edu/abs/2017arXiv171210246}
  {\bibfield  {journal} {\bibinfo  {journal} {ArXiv e-prints}\ } (\bibinfo
  {year} {2017})},\ \Eprint {http://arxiv.org/abs/1712.10246} {arXiv:1712.10246
  [nucl-th]} \BibitemShut {NoStop}%
\bibitem [{\citenamefont {Birkhan}\ \emph {et~al.}(2017)\citenamefont
  {Birkhan}, \citenamefont {Miorelli}, \citenamefont {Bacca}, \citenamefont
  {Bassauer}, \citenamefont {Bertulani}, \citenamefont {Hagen}, \citenamefont
  {Matsubara}, \citenamefont {von Neumann-Cosel}, \citenamefont {Papenbrock},
  \citenamefont {Pietralla}, \citenamefont {Ponomarev}, \citenamefont
  {Richter}, \citenamefont {Schwenk},\ and\ \citenamefont
  {Tamii}}]{birkhan2017}%
  \BibitemOpen
  \bibfield  {author} {\bibinfo {author} {\bibfnamefont {J.}~\bibnamefont
  {Birkhan}}, \bibinfo {author} {\bibfnamefont {M.}~\bibnamefont {Miorelli}},
  \bibinfo {author} {\bibfnamefont {S.}~\bibnamefont {Bacca}}, \bibinfo
  {author} {\bibfnamefont {S.}~\bibnamefont {Bassauer}}, \bibinfo {author}
  {\bibfnamefont {C.~A.}\ \bibnamefont {Bertulani}}, \bibinfo {author}
  {\bibfnamefont {G.}~\bibnamefont {Hagen}}, \bibinfo {author} {\bibfnamefont
  {H.}~\bibnamefont {Matsubara}}, \bibinfo {author} {\bibfnamefont
  {P.}~\bibnamefont {von Neumann-Cosel}}, \bibinfo {author} {\bibfnamefont
  {T.}~\bibnamefont {Papenbrock}}, \bibinfo {author} {\bibfnamefont
  {N.}~\bibnamefont {Pietralla}}, \bibinfo {author} {\bibfnamefont {V.~Yu.}\
  \bibnamefont {Ponomarev}}, \bibinfo {author} {\bibfnamefont {A.}~\bibnamefont
  {Richter}}, \bibinfo {author} {\bibfnamefont {A.}~\bibnamefont {Schwenk}}, \
  and\ \bibinfo {author} {\bibfnamefont {A.}~\bibnamefont {Tamii}},\ }\bibfield
   {title} {\enquote {\bibinfo {title} {Electric dipole polarizability of
  $^{48}\mathrm{Ca}$ and implications for the neutron skin},}\ }\href {\doibase
  10.1103/PhysRevLett.118.252501} {\bibfield  {journal} {\bibinfo  {journal}
  {Phys. Rev. Lett.}\ }\textbf {\bibinfo {volume} {118}},\ \bibinfo {pages}
  {252501} (\bibinfo {year} {2017})}\BibitemShut {NoStop}%
\bibitem [{\citenamefont {Wang}\ \emph {et~al.}(2012)\citenamefont {Wang},
  \citenamefont {Audi}, \citenamefont {Wapstra}, \citenamefont {Kondev},
  \citenamefont {MacCormick}, \citenamefont {Xu},\ and\ \citenamefont
  {Pfeiffer}}]{wang2012}%
  \BibitemOpen
  \bibfield  {author} {\bibinfo {author} {\bibfnamefont {M.}~\bibnamefont
  {Wang}}, \bibinfo {author} {\bibfnamefont {G.}~\bibnamefont {Audi}}, \bibinfo
  {author} {\bibfnamefont {A.~H.}\ \bibnamefont {Wapstra}}, \bibinfo {author}
  {\bibfnamefont {F.~G.}\ \bibnamefont {Kondev}}, \bibinfo {author}
  {\bibfnamefont {M.}~\bibnamefont {MacCormick}}, \bibinfo {author}
  {\bibfnamefont {X.}~\bibnamefont {Xu}}, \ and\ \bibinfo {author}
  {\bibfnamefont {B.}~\bibnamefont {Pfeiffer}},\ }\bibfield  {title} {\enquote
  {\bibinfo {title} {{The AME2012 atomic mass evaluation}},}\ }\href {\doibase
  10.1088/1674-1137/36/12/003} {\bibfield  {journal} {\bibinfo  {journal}
  {Chin. Phys. C}\ }\textbf {\bibinfo {volume} {36}},\ \bibinfo {pages} {1603}
  (\bibinfo {year} {2012})}\BibitemShut {NoStop}%
\bibitem [{\citenamefont {Giraud}(2008)}]{giraud2008}%
  \BibitemOpen
  \bibfield  {author} {\bibinfo {author} {\bibfnamefont {B.~G.}\ \bibnamefont
  {Giraud}},\ }\bibfield  {title} {\enquote {\bibinfo {title} {Density
  functionals in the laboratory frame},}\ }\href {\doibase
  10.1103/PhysRevC.77.014311} {\bibfield  {journal} {\bibinfo  {journal} {Phys.
  Rev. C}\ }\textbf {\bibinfo {volume} {77}},\ \bibinfo {pages} {014311}
  (\bibinfo {year} {2008})}\BibitemShut {NoStop}%
\bibitem [{\citenamefont {Hagen}\ \emph
  {et~al.}(2014{\natexlab{b}})\citenamefont {Hagen}, \citenamefont
  {Papenbrock}, \citenamefont {Ekstr\"om}, \citenamefont {Wendt}, \citenamefont
  {Baardsen}, \citenamefont {Gandolfi}, \citenamefont {Hjorth-Jensen},\ and\
  \citenamefont {Horowitz}}]{hagen2013b}%
  \BibitemOpen
  \bibfield  {author} {\bibinfo {author} {\bibfnamefont {G.}~\bibnamefont
  {Hagen}}, \bibinfo {author} {\bibfnamefont {T.}~\bibnamefont {Papenbrock}},
  \bibinfo {author} {\bibfnamefont {A.}~\bibnamefont {Ekstr\"om}}, \bibinfo
  {author} {\bibfnamefont {K.~A.}\ \bibnamefont {Wendt}}, \bibinfo {author}
  {\bibfnamefont {G.}~\bibnamefont {Baardsen}}, \bibinfo {author}
  {\bibfnamefont {S.}~\bibnamefont {Gandolfi}}, \bibinfo {author}
  {\bibfnamefont {M.}~\bibnamefont {Hjorth-Jensen}}, \ and\ \bibinfo {author}
  {\bibfnamefont {C.~J.}\ \bibnamefont {Horowitz}},\ }\bibfield  {title}
  {\enquote {\bibinfo {title} {Coupled-cluster calculations of nucleonic
  matter},}\ }\href {\doibase 10.1103/PhysRevC.89.014319} {\bibfield  {journal}
  {\bibinfo  {journal} {Phys. Rev. C}\ }\textbf {\bibinfo {volume} {89}},\
  \bibinfo {pages} {014319} (\bibinfo {year} {2014}{\natexlab{b}})}\BibitemShut
  {NoStop}%
\bibitem [{\citenamefont {Gandolfi}\ \emph {et~al.}(2009)\citenamefont
  {Gandolfi}, \citenamefont {Illarionov}, \citenamefont {Schmidt},
  \citenamefont {Pederiva},\ and\ \citenamefont {Fantoni}}]{gandolfi2009}%
  \BibitemOpen
  \bibfield  {author} {\bibinfo {author} {\bibfnamefont {S.}~\bibnamefont
  {Gandolfi}}, \bibinfo {author} {\bibfnamefont {A.~Yu.}\ \bibnamefont
  {Illarionov}}, \bibinfo {author} {\bibfnamefont {K.~E.}\ \bibnamefont
  {Schmidt}}, \bibinfo {author} {\bibfnamefont {F.}~\bibnamefont {Pederiva}}, \
  and\ \bibinfo {author} {\bibfnamefont {S.}~\bibnamefont {Fantoni}},\
  }\bibfield  {title} {\enquote {\bibinfo {title} {Quantum monte carlo
  calculation of the equation of state of neutron matter},}\ }\href {\doibase
  10.1103/PhysRevC.79.054005} {\bibfield  {journal} {\bibinfo  {journal} {Phys.
  Rev. C}\ }\textbf {\bibinfo {volume} {79}},\ \bibinfo {pages} {054005}
  (\bibinfo {year} {2009})}\BibitemShut {NoStop}%
\bibitem [{\citenamefont {Tsang}\ \emph {et~al.}(2012)\citenamefont {Tsang},
  \citenamefont {Stone}, \citenamefont {Camera}, \citenamefont {Danielewicz},
  \citenamefont {Gandolfi}, \citenamefont {Hebeler}, \citenamefont {Horowitz},
  \citenamefont {Lee}, \citenamefont {Lynch}, \citenamefont {Kohley},
  \citenamefont {Lemmon}, \citenamefont {M\"oller}, \citenamefont {Murakami},
  \citenamefont {Riordan}, \citenamefont {Roca-Maza}, \citenamefont
  {Sammarruca}, \citenamefont {Steiner}, \citenamefont {Vida\~na},\ and\
  \citenamefont {Yennello}}]{tsang2012}%
  \BibitemOpen
  \bibfield  {author} {\bibinfo {author} {\bibfnamefont {M.~B.}\ \bibnamefont
  {Tsang}}, \bibinfo {author} {\bibfnamefont {J.~R.}\ \bibnamefont {Stone}},
  \bibinfo {author} {\bibfnamefont {F.}~\bibnamefont {Camera}}, \bibinfo
  {author} {\bibfnamefont {P.}~\bibnamefont {Danielewicz}}, \bibinfo {author}
  {\bibfnamefont {S.}~\bibnamefont {Gandolfi}}, \bibinfo {author}
  {\bibfnamefont {K.}~\bibnamefont {Hebeler}}, \bibinfo {author} {\bibfnamefont
  {C.~J.}\ \bibnamefont {Horowitz}}, \bibinfo {author} {\bibfnamefont {Jenny}\
  \bibnamefont {Lee}}, \bibinfo {author} {\bibfnamefont {W.~G.}\ \bibnamefont
  {Lynch}}, \bibinfo {author} {\bibfnamefont {Z.}~\bibnamefont {Kohley}},
  \bibinfo {author} {\bibfnamefont {R.}~\bibnamefont {Lemmon}}, \bibinfo
  {author} {\bibfnamefont {P.}~\bibnamefont {M\"oller}}, \bibinfo {author}
  {\bibfnamefont {T.}~\bibnamefont {Murakami}}, \bibinfo {author}
  {\bibfnamefont {S.}~\bibnamefont {Riordan}}, \bibinfo {author} {\bibfnamefont
  {X.}~\bibnamefont {Roca-Maza}}, \bibinfo {author} {\bibfnamefont
  {F.}~\bibnamefont {Sammarruca}}, \bibinfo {author} {\bibfnamefont {A.~W.}\
  \bibnamefont {Steiner}}, \bibinfo {author} {\bibfnamefont {I.}~\bibnamefont
  {Vida\~na}}, \ and\ \bibinfo {author} {\bibfnamefont {S.~J.}\ \bibnamefont
  {Yennello}},\ }\bibfield  {title} {\enquote {\bibinfo {title} {Constraints on
  the symmetry energy and neutron skins from experiments and theory},}\ }\href
  {\doibase 10.1103/PhysRevC.86.015803} {\bibfield  {journal} {\bibinfo
  {journal} {Phys. Rev. C}\ }\textbf {\bibinfo {volume} {86}},\ \bibinfo
  {pages} {015803} (\bibinfo {year} {2012})}\BibitemShut {NoStop}%
\bibitem [{\citenamefont {{Tews}}\ \emph {et~al.}(2016)\citenamefont {{Tews}},
  \citenamefont {{Lattimer}}, \citenamefont {{Ohnishi}},\ and\ \citenamefont
  {{Kolomeitsev}}}]{tews2016}%
  \BibitemOpen
  \bibfield  {author} {\bibinfo {author} {\bibfnamefont {I.}~\bibnamefont
  {{Tews}}}, \bibinfo {author} {\bibfnamefont {J.~M.}\ \bibnamefont
  {{Lattimer}}}, \bibinfo {author} {\bibfnamefont {A.}~\bibnamefont
  {{Ohnishi}}}, \ and\ \bibinfo {author} {\bibfnamefont {E.~E.}\ \bibnamefont
  {{Kolomeitsev}}},\ }\bibfield  {title} {\enquote {\bibinfo {title} {{Symmetry
  Parameter Constraints From A Lower Bound On The Neutron-Matter Energy}},}\
  }\href@noop {} {\bibfield  {journal} {\bibinfo  {journal} {ArXiv e-prints}\ }
  (\bibinfo {year} {2016})},\ \Eprint {http://arxiv.org/abs/1611.07133}
  {arXiv:1611.07133 [nucl-th]} \BibitemShut {NoStop}%
\end{thebibliography}%

\end{document}